\documentclass{article}

\usepackage{arxiv}

\usepackage[utf8]{inputenc} 
\usepackage[T1]{fontenc}    
\usepackage{hyperref}       
\usepackage{url}            
\usepackage{booktabs}       
\usepackage{amsfonts}       
\usepackage{nicefrac}       
\usepackage{microtype}      
\usepackage{lipsum}
\usepackage{graphicx}
\usepackage{amssymb}
\usepackage{amsmath}
\usepackage{enumitem}
\usepackage{subcaption}
\usepackage{placeins}
\usepackage{matlab-prettifier}
\usepackage{xcolor}
\usepackage[style=nature]{biblatex}
\AtEveryBibitem{\clearfield{isbn}}
\AtEveryCitekey{\clearfield{isbn}}
\AtEveryBibitem{\clearfield{issn}}
\AtEveryCitekey{\clearfield{issn}}
\AtEveryBibitem{\clearfield{doi}}
\AtEveryCitekey{\clearfield{doi}}
\AtEveryBibitem{\clearfield{url}}
\AtEveryCitekey{\clearfield{url}}
\AtEveryBibitem{\clearfield{urlyear}}
\AtEveryCitekey{\clearfield{urldate}}
\AtEveryBibitem{\clearfield{note}}
\AtEveryCitekey{\clearfield{note}}
\AtEveryBibitem{\clearfield{month}}
\AtEveryBibitem{\clearfield{day}}
\AtEveryBibitem{\clearfield{pagetotal}}
\graphicspath{ {./images/} }
\title{The Complex Interplay Between Risk Tolerance and the Spread of Infectious Diseases}

\author{
Maximilian Nguyen \\
Lewis-Sigler Institute\\
Princeton University\\
Princeton, NJ 08544 \\
\texttt{mmnguyen@princeton.edu} \And
Ari Freedman\\
Department of Ecology and Evolutionary Biology\\
Princeton University\\
Princeton, NJ, USA\\
\texttt{arisf@princeton.edu} \And
Matthew Cheung\\
Program in Applied and Computational Mathematics\\
Princeton University\\
Princeton, NJ, USA\\
\texttt{matthew.cheung@princeton.edu}  \And
Chadi Saad-Roy\\
Miller Institute for Basic Research in Science\\ Department of Integrative Biology\\
University of California, Berkeley\\
Berkeley, CA, USA\\
\texttt{csaadroy@berkeley.edu} \And
Baltazar Espinoza\\
Biocomplexity Institute\\
University of Virginia\\
Charlottesville, VA, USA\\
\texttt{baltazar.espinoza@virginia.edu} \And
Bryan Grenfell\\
Department of Ecology and Evolutionary Biology\\
Princeton University\\
Princeton, NJ, USA\\
\texttt{grenfell@princeton.edu}  \And
Simon Levin\\
Department of Ecology and Evolutionary Biology\\
Princeton University\\
Princeton, NJ, USA\\
\texttt{slevin@princeton.edu}
}

\begin{document}
\maketitle




\begin{abstract}
Risk-driven behavior provides a feedback mechanism through which individuals both shape and are collectively affected by an epidemic. We introduce a general and flexible compartmental model to study the effect of heterogeneity in the population with regards to risk tolerance. The interplay between behavior and epidemiology leads to a rich set of possible epidemic dynamics. Depending on the behavioral composition of the population, we find that increasing heterogeneity in risk tolerance can either increase or decrease the epidemic size. We find that multiple waves of infection can arise due to the interplay between transmission and behavior, even without the replenishment of susceptibles. We find that increasing protective mechanisms such as the effectiveness of interventions, the number of risk-averse people in the population, and the duration of intervention usage reduces the epidemic overshoot. When the protection is pushed past a critical threshold, the epidemic dynamics enter an underdamped regime where the epidemic size exactly equals the herd immunity threshold. Lastly, we can find regimes where epidemic size does not monotonically decrease with a population that becomes increasingly risk-averse.
\end{abstract}

\section*{Introduction}
Recent outbreaks such as the COVID-19 pandemic, the 2014 Ebola outbreak, the 2009 influenza A (H1N1) pandemic, and the 2002 SARS epidemic brought to light many of the challenges of mounting an effective and unified epidemic response in a country as large and as diverse as the United States. Particularly during the COVID-19 pandemic, people were split in opinion on questions such as the origin of the virus \cite{maxmen_covid_2021}, whether they would social distance or wear a mask \cite{betsch_social_2020, fischer_mask_2021, yang_sociocultural_2022}, or whether the country should even have a pandemic response at all \cite{morens_concept_2022}. As time progressed, the situation became more dire and the death toll accumulated. People then had a new battery of questions to address, such as whether or not they would adhere to mandatory lock-downs \cite{wong_paradox_2020, brzezinski_science_2021, kleitman_comply_2021} or whether they felt comfortable using the new mRNA vaccines \cite{machingaidze_understanding_2021, fedele_covid-19_2021}. Compounding the issue were the multiple streams of information and potential misinformation spread through social media and other channels \cite{gallotti_assessing_2020, loomba_measuring_2021, offer-westort_battling_2024, towers_mass_2015}. People's stances on the questions and issues were diverse, arising from the milieu of differences in culture, geography, scientific education, sources of information, political leanings, and individual identity \cite{ murray_chapter_2016, kramer_infection_2021, lu_collectivism_2021}. 

Taken altogether, these differences within the population reflect a spectrum of people's risk tolerances to a circulating infectious disease. For any given intervention, such as social distancing, wearing a mask, or taking a vaccine, each person in the population falls somewhere on a spectrum of willingness to adopt the intervention. Given a threat level of an infectious disease in the population, some people will readily wear masks, whereas other people will refuse to. 

In this study, we aim to analyze the impact of heterogeneity in risk tolerance and the resulting behavioral response on the dynamics of epidemics. We seek to add to a burgeoning literature on the impact of human behavior in epidemic response \cite{bansal_when_2007, funk_modelling_2010, fenichel_adaptive_2011, edmunds_evaluating_1999, weitz_awareness-driven_2020, wagner_economic_2020, tyson_timing_2020, espinoza_asymptomatic_2021, wagner_modelling_2022, espinoza_heterogeneous_2022, qiu_understanding_2022, traulsen_individual_2023, saad-roy_dynamics_2023, smith_covid-19_2023}, which the recent pandemic highlighted as an area for further exploration in preparation for the next large scale global health crisis \cite{morse_prediction_2012, osterholm_preparing_2020, bergstrom_human_2024}. To study the impact of heterogeneity in risk tolerance on epidemic dynamics, we introduce a simple and flexible modeling framework based on ordinary differential equations that can be used for different interventions and an arbitrary partitioning of the population with regard to risk tolerance and behavioral responses. We will examine and discuss potential interesting outcomes that can arise from coupling individual-level preferences and population-level epidemiology. 


\section*{Results}
\subsection*{Model of Adaptive Intervention Usage under Heterogeneous Risk Tolerance}
Here we assume people's risk aversion manifests as the rate at which they adopt individual interventions in response to an infectious disease outbreak. The intuition underlying this paradigm is that more risk-averse individuals are more sensitive to becoming sick and thus will adopt interventions at a faster rate than more risk-tolerant people. We consider the following SPIR compartmental model of a population with $n$ differing levels of risk tolerance (\ref{eqn:modelfirst}-\ref{eqn:modellast}). This model features four types of classes: unprotected susceptible ($S$), protected susceptible ($P$), infectious ($I$), and recovered with permanent immunity ($R$). Since there are $n$ differing levels of risk tolerance, we subdivide the susceptible population into $n$ discrete groups indexed by $i$, where $i\in \{1, 2, ..., n\}$. Each tolerance level is characterized by an intervention adoption rate parameter ($\lambda_i$) and an intervention relaxation rate parameter ($\delta_i$). Transitions of susceptibles between their unprotected class ($S_i$) and their corresponding protected class ($P_i$) are governed by 
 the corresponding parameters of the same index ($\lambda_i,\delta_i$). Overall, the system is governed by $3+2n$ parameters: a transmission rate parameter ($\beta$), a recovery rate parameter ($\gamma$), an intervention effectiveness parameter ($\epsilon$), and an intervention adoption rate ($\lambda_i$) and intervention relaxation rate ($\delta_i$) for each tolerance level.

\begin{align}
\frac{dS_i}{dt} &= -\beta S_i I -\lambda_i S_i I + \delta_i P_i \label{eqn:modelfirst}\\
\frac{dP_i}{dt} &= -(1-\epsilon)\beta P_i I - \delta_i P_i + \lambda_i S_i I \\
\frac{dI}{dt} &= - \gamma I+\sum_{i=1}^n (\beta S_i I + (1-\epsilon)\beta P_i I)\\
\frac{dR}{dt} &= \gamma I \label{eqn:modellast}
\end{align}

The transition from the unprotected susceptible state to the protected susceptible state represents individuals implementing an intervention that confers them protection against disease transmission from an infected individual. The rate at which intervention adoption occurs may be driven by individuals considering information such as the epidemic incidence rate (e.g. cases per day), the total number of infected individuals in the population (e.g. total number of active cases), and mortality rate (e.g. deaths per day) \cite{weitz_awareness-driven_2020}. Here we assume that individuals have knowledge about the total number of infected individuals ($I$) and respond accordingly. Parameterizing each person’s individual risk tolerance by $\lambda_i$, we assume each individual person adopts an intervention at a rate $\lambda_i I$. Then, if there are $S_i$ number of people that behave exactly the same (i.e. have the same level of risk-aversion), then at the population scale there is a collective adoption rate of $\lambda_i S_i I$. The same reasoning holds for each of the $n$ tolerance levels. We also consider a model where the adoption rate is driven by individuals reacting to the incidence rate (Supplemental Materials); while this produces a more complex mathematical model, the results are qualitatively similar.

The effectiveness of the intervention being used is captured by the parameter $\epsilon$, which linearly scales down the transmission rate between infected and protected susceptibles. In the limit of $\epsilon=1$, the intervention is perfectly effective and protected individuals cannot become infected. In the limit of $\epsilon=0$ then the intervention is completely ineffective, which reduces the model to an SIR model without interventions. For simplicity, we assume each epidemic features only a single type of intervention (whether that be masking, social distancing, vaccines, etc.) and that the effectiveness of an intervention is identical across the population. In reality, multiple interventions may be available concurrently, which would drive additional variation in behavior due to differences in risk sensitivity across the population. 

This model allows for protected individuals to relax their usage of interventions, becoming unprotected in the process. Here, individuals in the protected class can relax back to the unprotected class through two means, either through a rate that is dependent on the quantity of infections present or through a rate that is independent of the number of infections. The infection-dependent rate is implicitly captured through the $\lambda SI$ term, which can be thought of a net rate that can be further decomposed into adoption and relaxation rate terms (i.e. $\lambda SI = \lambda _{adoption}SI-\lambda _{relaxation}SI$). Here we have assumed the adoption term to always be greater than the relaxation term, otherwise individuals would never adopt an intervention. The infection-independent rate is governed by the intervention relaxation rate parameter ($\delta_i$) for each tolerance level. In the limit of $\delta_i=0$, an intervention is irreversible, which would represent an intervention such as vaccines with permanent immunity. When $\delta_i$ is non-zero, individuals are using interventions such as masking or social distancing. The infection-independent rate is motivated by factors such as psychological fatigue of social distancing \cite{franzen_fatigue_2021, jorgensen_pandemic_2022} and physical discomfort with wearing masks \cite{cheok_appropriate_2021}. In general, we will consider the regime where the relaxation rate $\delta$ is of comparable scale or smaller than the transmission scale (i.e. $\delta \leq \beta$). This reflects intuition that people are likely to continue to protect themselves with interventions even beyond an initial outbreak \cite{barak_experience_2022}. 

\begin{figure}[ht]
\centering
\includegraphics[scale=0.4]{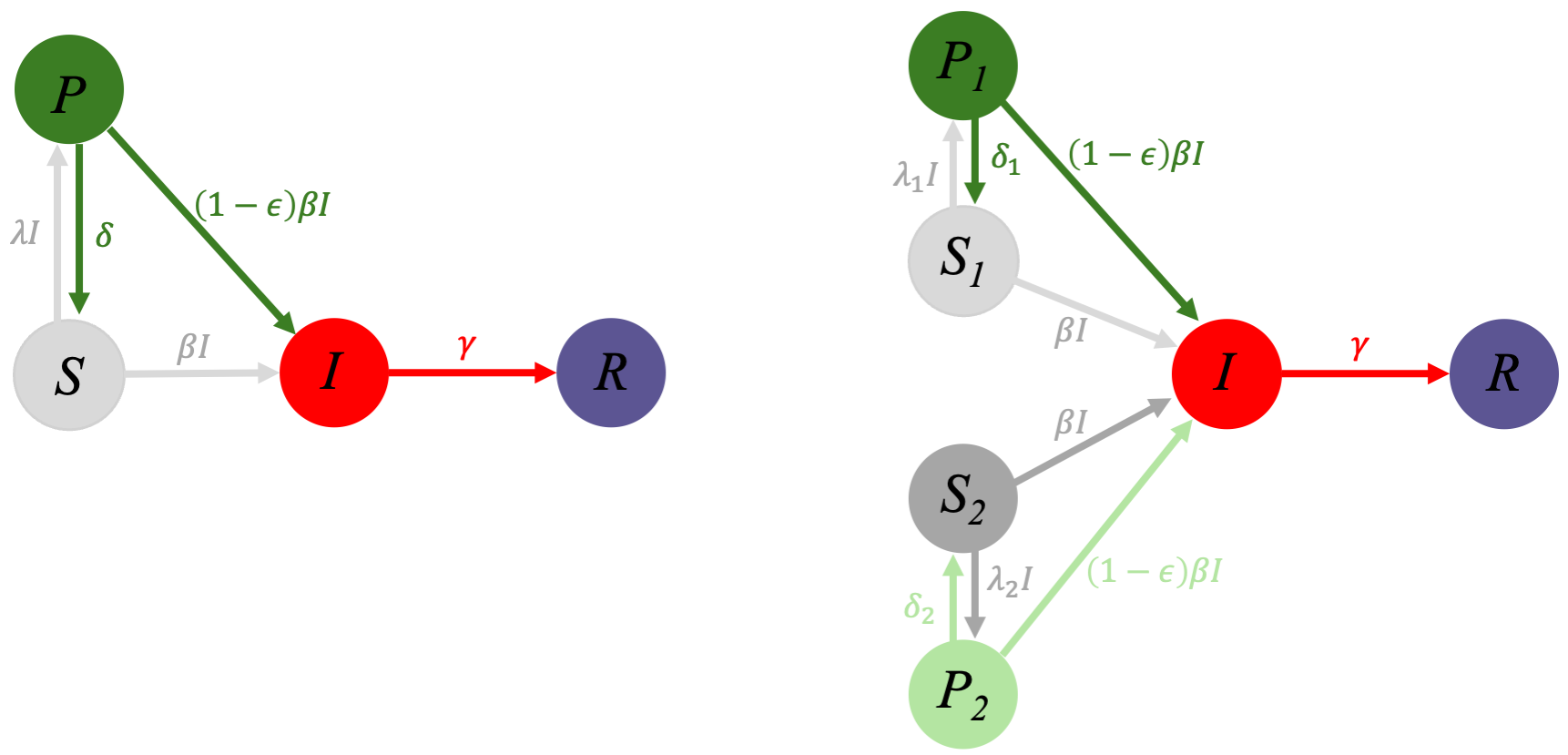}
\caption{Flow diagram for an SIR model with adaptive interventions for either (a) a population with homogeneous risk tolerance or (b) a heterogeneous population with two different levels of risk tolerance. Susceptible individuals can access a more protected susceptible state through usage of interventions. The transition rate to the protected state depends on the incidence level. The protected state offers a $1-\epsilon$ reduction in transmission rate over the normal susceptible state.} \label{fig:flowchart}
\end{figure}

For simplicity, we consider the model for the case when $n=1$ and $n=2$. A schematic for these two cases is shown in Figure \ref{fig:flowchart}. However, the framework is general and can be extended to any discrete number of groups.

For convention, when there are two susceptible classes, we assume the first susceptible class ($S_1$) has a lower risk tolerance for becoming infected (i.e. more risk-averse). As a result, these individuals more readily adopt the intervention (i.e. $\lambda_1 >\lambda_2$), making individuals in this class transition more rapidly to the protected susceptible state ($P_1$). The second susceptible class ($S_2$) is more risk tolerant (i.e. more risk-taking), and thus is less eager to use the intervention, making individuals in this class transition more slowly to their protected susceptible state ($P_2$).

\subsubsection*{Adaptive Adoption of Interventions Can Produce Damped Oscillations}
\begin{figure}[ht]
\centering
\includegraphics[scale=0.5]{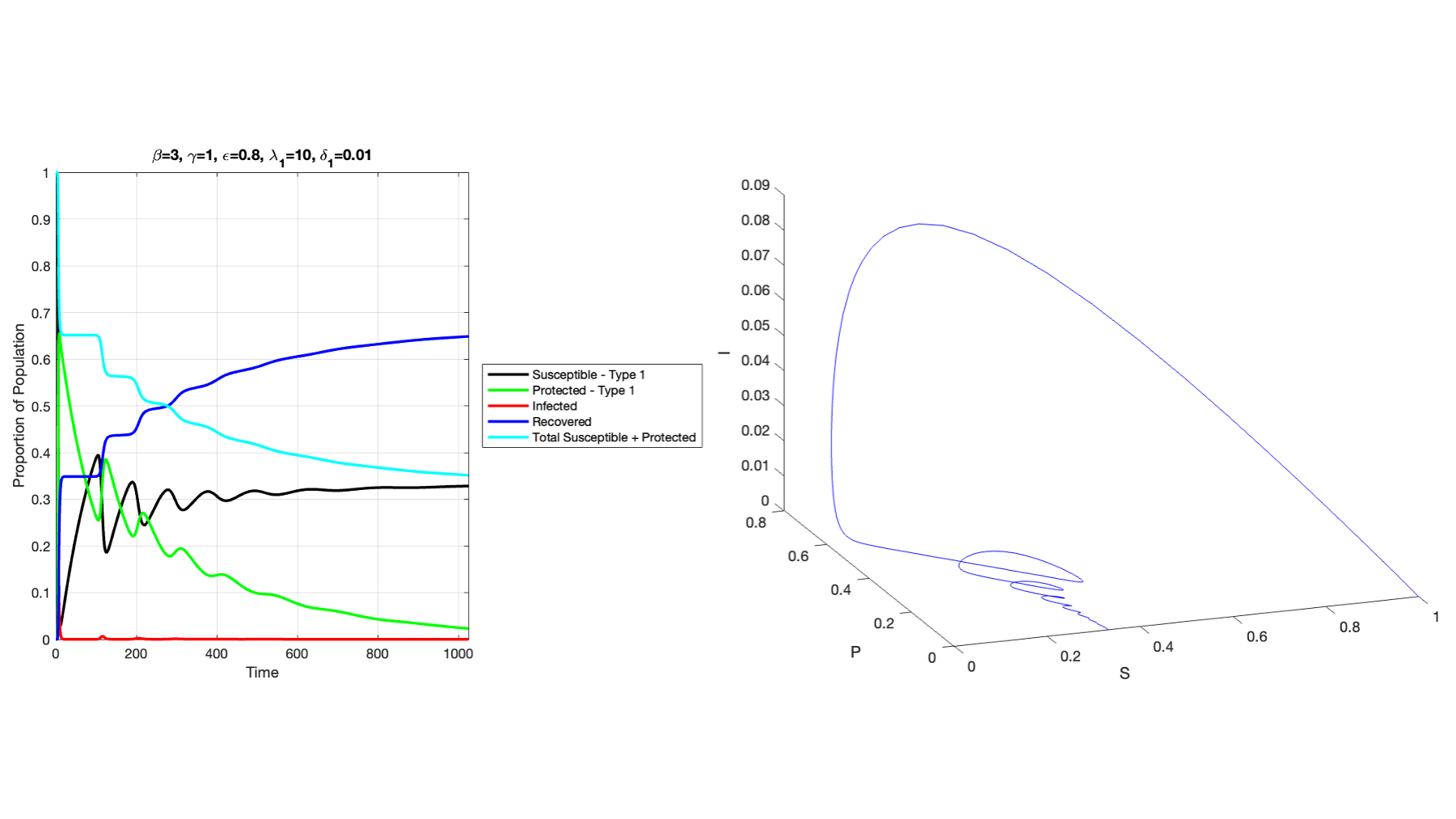}
\caption{Time series for population with homogeneous risk tolerance and adaptive intervention usage and the corresponding phase space trajectory indicate the presence of damped oscillations.} \label{fig:sensitivity}
\end{figure}

The coupling of intervention usage to the incidence rate and the resulting adaptive changes enables the epidemic dynamics to display a much richer set of behavior over the simple SIR model. From Figure \ref{fig:sensitivity}, we see this particular set of conditions can deterministically produce multiple waves of infection, even when vital dynamics (i.e. birth and death processes) are not considered.

Evidence for cycling of individuals between using interventions and not using interventions during the COVID pandemic can be seen in longitudinal usage \cite{mathieu_coronavirus_2020, hale_global_2021, rader_mask-wearing_2021, salomon_us_2021} data. The possibility for these oscillations highlight the intimate connection between individual human behavior and intervention usage in shaping the dynamics of epidemics, while also be affected by the collective decision of everyone in the population. The coupling of behavior and epidemiology here provides a feedback mechanism where an increasing incidence rate prompts more individuals to adopt an intervention, which lowers the overall incidence rate; however, as the epidemic wanes and factors such as fatigue or discomfort set in, people begin dropping their usage of interventions, which may eventually lead to another wave of outbreaks if enough people become unprotected while infected individuals still remain, and then the cycle can be repeated. 

\subsubsection*{Protective Mechanisms Saturate in Underdamped Regime that Eliminates Epidemic Overshoot}
One might have the intuition that having more people that will more readily adopt an intervention (i.e. mask, social distance, or vaccinate) or increasing the effectiveness of the intervention in reducing transmission will further decrease the size of the epidemic. While we find this intuition to be mostly correct, we unexpectedly find that the protection conferred by either of these mechanisms can saturate once a critical parameter threshold has been passed. 

\begin{figure}[ht]
\centering
\includegraphics[scale=0.2]{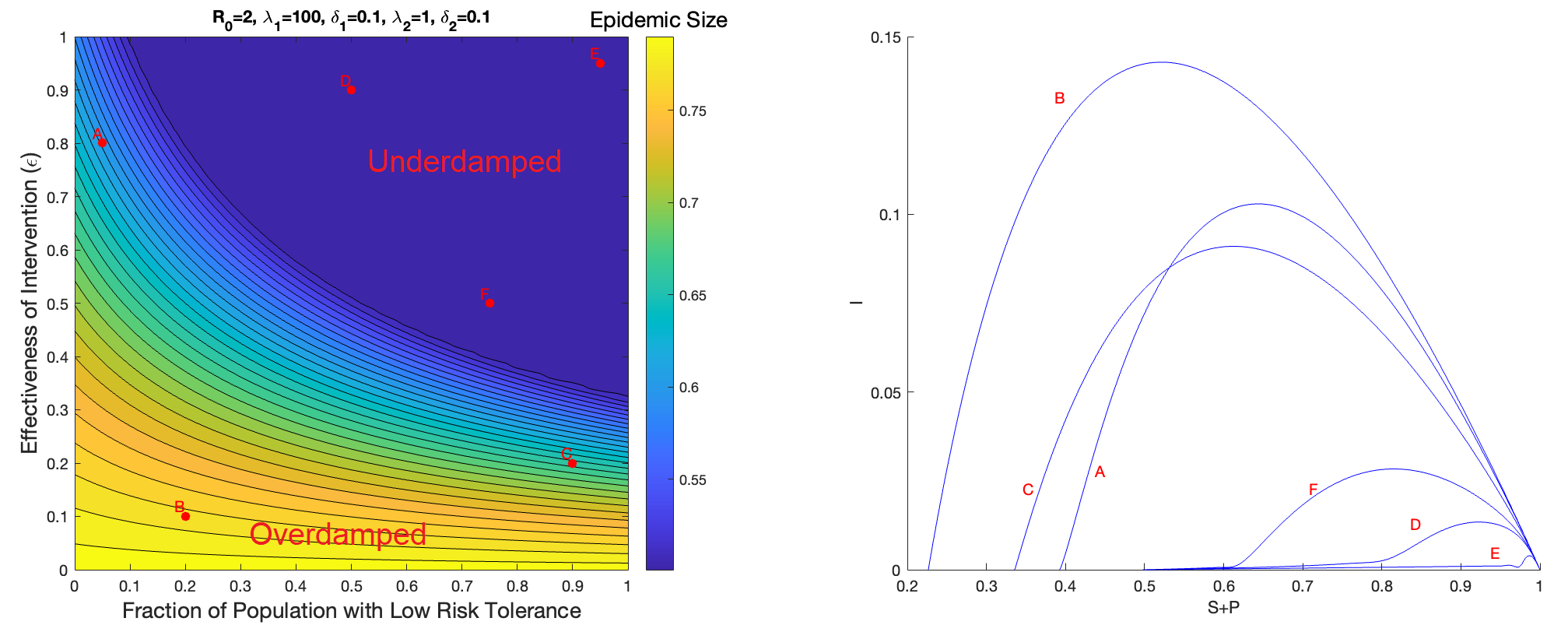}
\caption{Left. Epidemic size as a function of varying the fraction of the population that are low-risk tolerance (i.e. those with higher $\lambda$). Right. Corresponding orbits in the I versus S+P plane for the sampled points in parameter space.} \label{fig:saturation}
\end{figure}

In Figure \ref{fig:saturation}$, left$, we see that increasing the effectiveness of the intervention or increasing the fraction of the population that are risk-averse monotonically decreases the epidemic size. However, in the dark blue region (which we will refer to as the underdamped regime) where both protection mechanisms are at their highest, we see no further reduction in the epidemic size. This regime corresponds to an epidemic where the epidemic size exactly equals the herd immunity threshold.

The orbits of the dynamics from different areas of this parameter space are shown in Figure \ref{fig:saturation}$, right$. We see that even though the final epidemic size is the same throughout the underdamped region, the trajectories to reach the same final epidemic size can look qualitatively different.

Figures \ref{fig:saturationFixedX}-\ref{fig:saturationFixedEpsilon} show a larger sampling of trajectories if one fixes either the fraction of the population with low risk tolerance or the intervention effectiveness respectively. It becomes clear that at the border of the underdamped region, we can see a clear change in the qualitative behavior of the trajectories as the threshold is crossed. Under some assumptions, one can prove that that the epidemic overshoot is eliminated in the underdamped regime (Supplemental Information).

However, we should make the point that this is not evidence that highly effective interventions are a waste or that the overall population should tolerate risky behavior. As this is a model with a large parameter space, we cannot visualize all of it. If we could, we would find many parameter regimes where the protection mechanisms never reach a critical threshold, which implies the conventional intuition of increasing intervention effectiveness and having more risk-averse people always being beneficial applies.

\subsubsection*{The Threshold to the Underdamped Regime is Reduced when Intervention Usage is Prolonged}
\begin{figure*}[t!]
    \centering
    \begin{subfigure}[t]{0.5\textwidth}
        \centering
        \includegraphics[scale=.4]{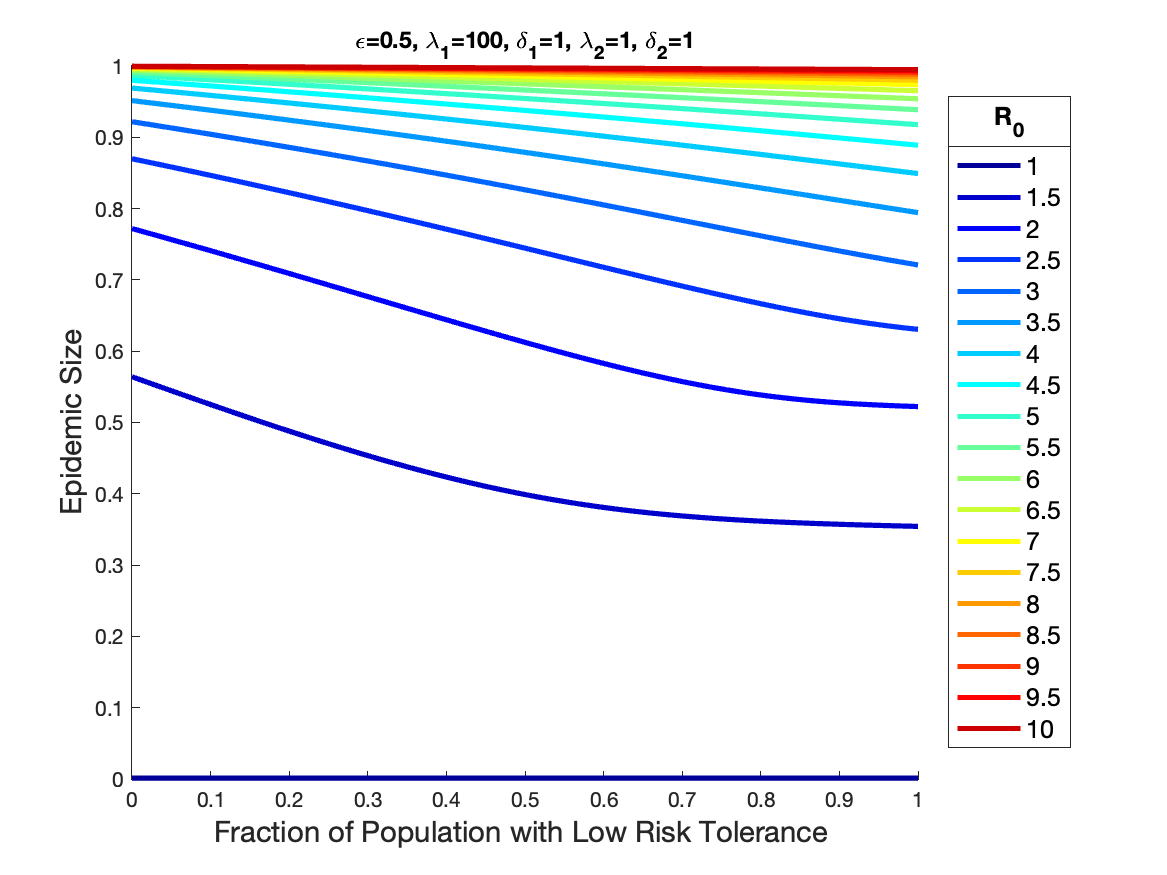}
        \caption{$\epsilon=0.5, \delta_1=\delta_2  = 1$}
    \end{subfigure}%
    ~ 
    \begin{subfigure}[t]{0.5\textwidth}
        \centering
        \includegraphics[scale=.4]{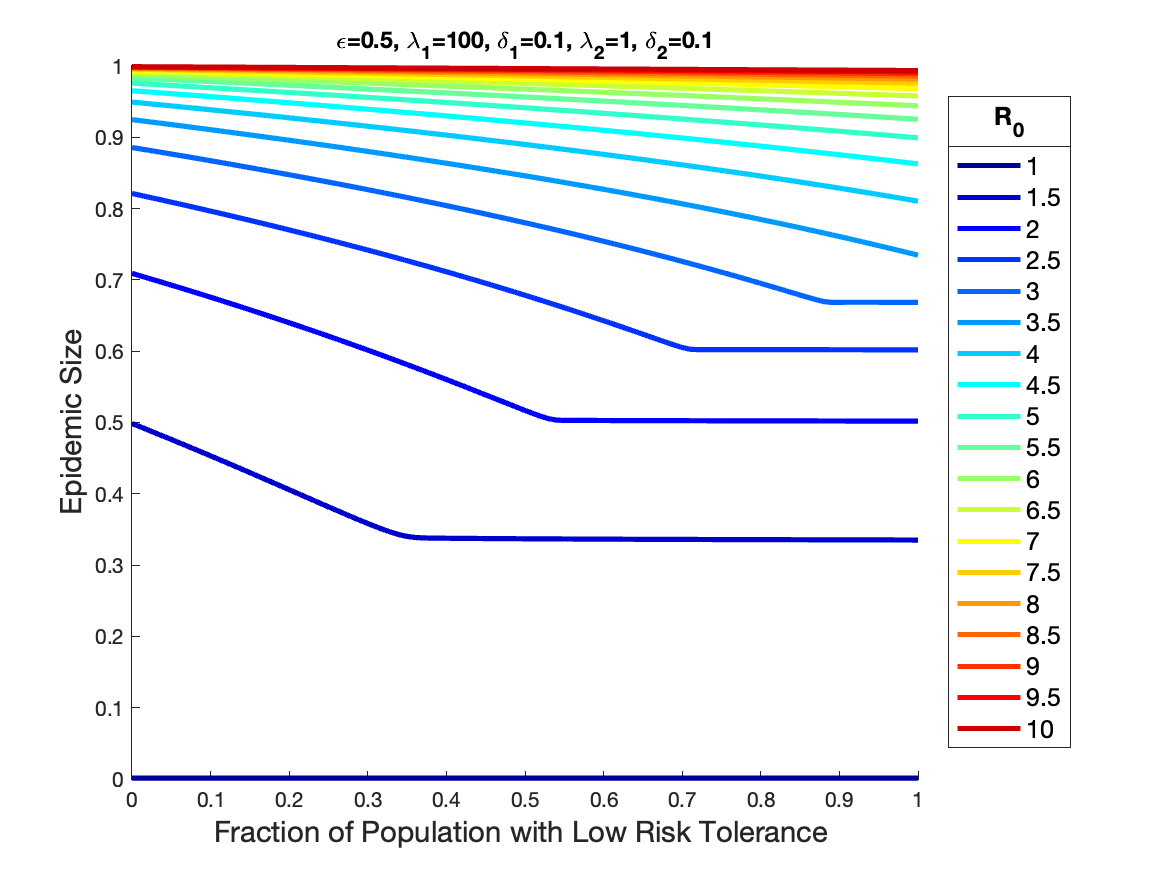}
        \caption{$\epsilon=0.5, \delta_1=\delta_2  = 0.1$}
    \end{subfigure}%
     

    \begin{subfigure}[t]{0.5\textwidth}
        \centering
        \includegraphics[scale=.4]{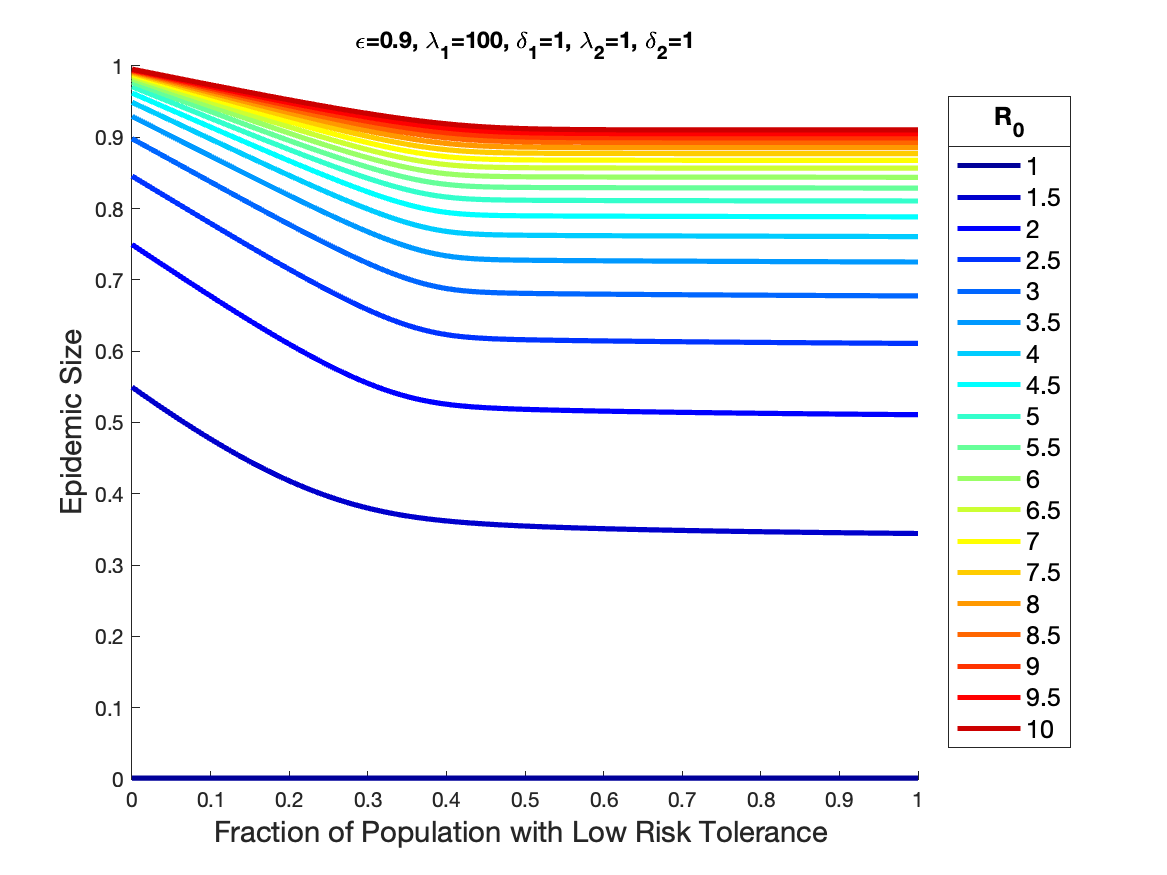}
        \caption{$\epsilon=0.9, \delta_1=\delta_2  = 1$}
    \end{subfigure}%
    ~
    \begin{subfigure}[t]{0.5\textwidth}
        \centering
        \includegraphics[scale=.4]{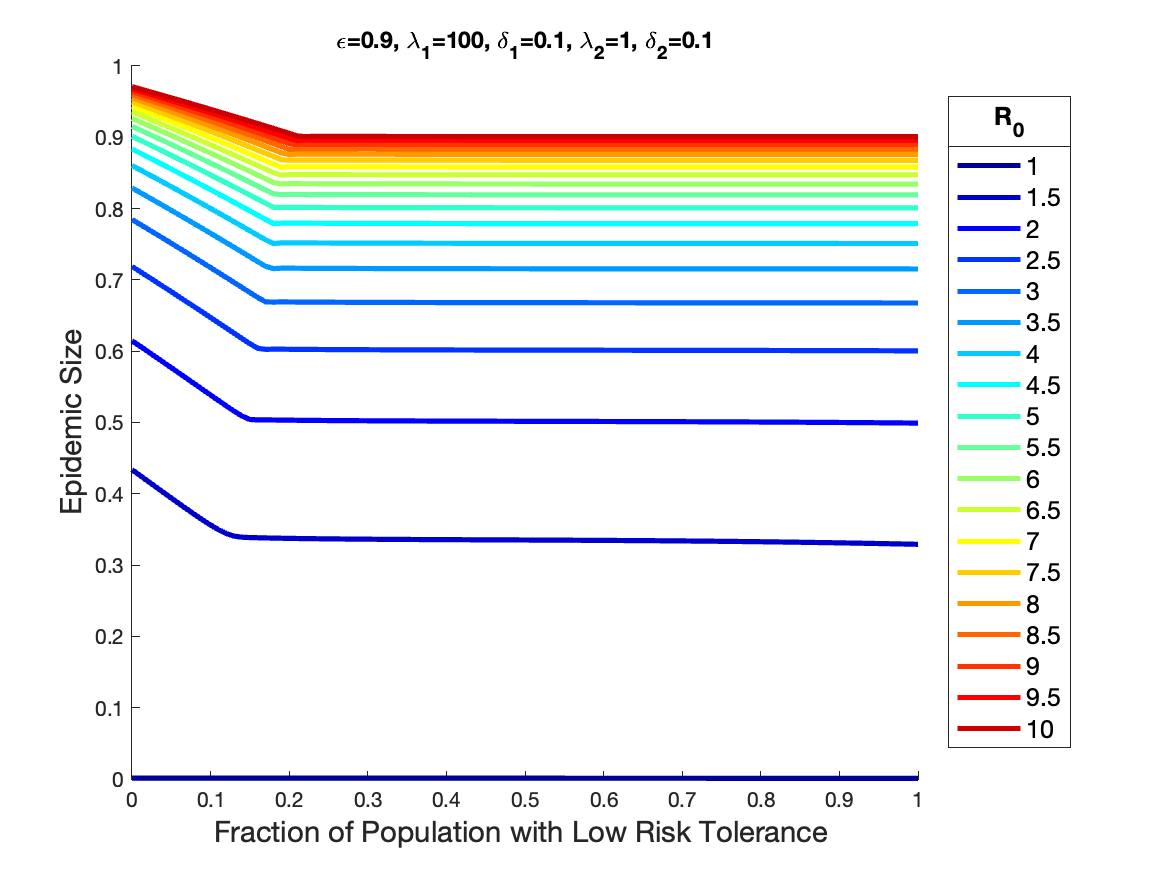}
        \caption{$\epsilon=0.9, \delta_1=\delta_2 = 0.1$}
    \end{subfigure}
    \caption{Final epidemic size versus fraction of population that are risk-averse ($S_1$). Simulations in the left column have a higher $\delta$ than simulations in the right column. The other simulation parameters and initial conditions are $\lambda_1 = 100, \lambda_2 = 1, I = 10^{-7}, P_1=P_2=0, R = 0$.} \label{fig:sharpness}
\end{figure*}

The transition to the underdamped regime is more easily accessed when the usage of interventions is prolonged (or equivalently when the rate at which protected individuals relax back into the regular susceptibility classes decreases). Consider the following scenario which is identical to the previous setup, except now the intervention reversion rate ($\delta P_i$) has been reduced by an order of magnitude (Figure \ref{fig:sharpness}). This corresponds to a scenario where people continue to use the intervention (i.e. such as wearing masks or social distancing) on a timescale significantly longer than the transmission timescale ($\delta >> \beta$).

This suggests that increasing the timescale at which individuals continue to use interventions decreases the number of risk-averse individuals needed to achieve the same epidemic size. This is reflected in the horizontal shift of the transition region to the left when comparing figures in the left column and the right column (Figure \ref{fig:sharpness}).

\FloatBarrier
\subsubsection*{Increasing Heterogeneity in Risk Tolerance can Either Increase or Decrease the Epidemic Size}
The literature generally suggests that increasing heterogeneity in the population through increasing the variation in their contact patterns, age, etc. results in a reduction in the epidemic size \cite{miller_epidemic_2007, britton_mathematical_2020, gomes_individual_2022, allard_role_2023}. 

\begin{figure}[ht]
\centering
\includegraphics[scale=0.17]{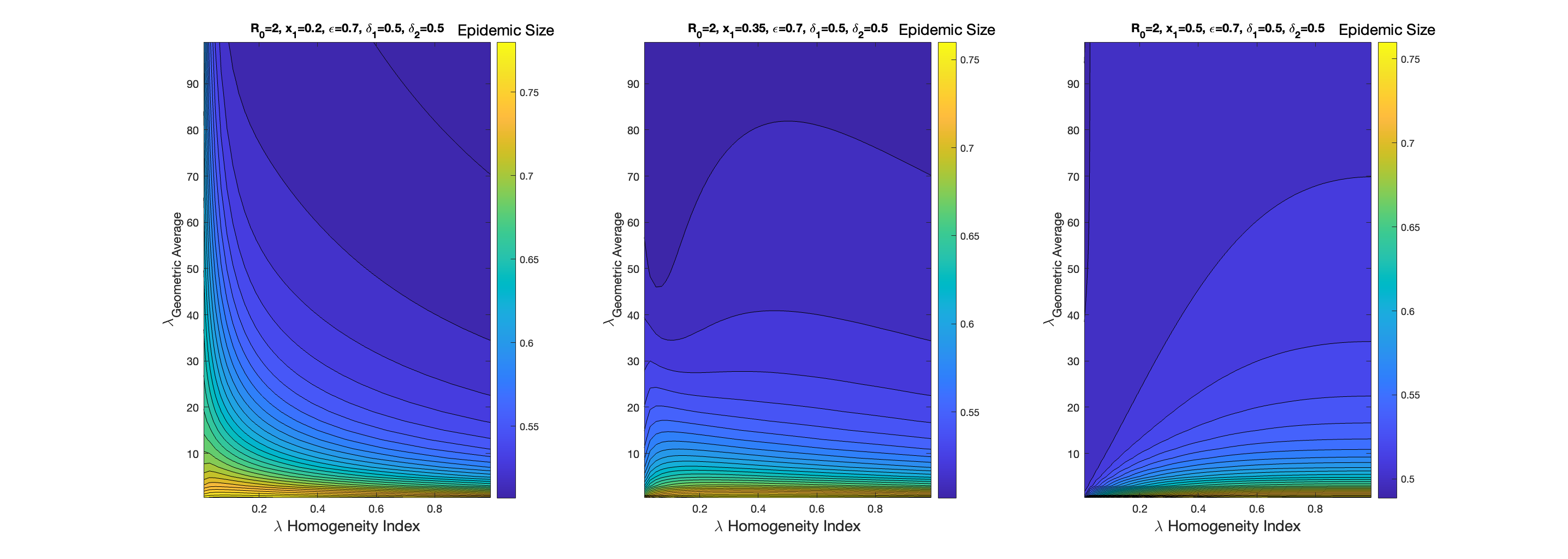}
\caption{Epidemic size under differing levels of heterogeneity in the adoption rate for interventions. The mean adoption rate of the two groups (i.e. geometric average of $\lambda_1, \lambda_2$) is compared to the difference in the two adoption rates as parameterized by a homogeneity index (see Methods for definition). $Left$ is when the fraction of the population with low risk tolerance ($x_1$) is 0.2, $center$ is when $x_1=0.35$, $right$ is when $x_1=0.5$.} \label{fig:heterogeneity}
\end{figure}

We find in this model of heterogeneous behavior that it is possible to switch from a regime where increasing the heterogeneity in risk tolerance results in a decrease in epidemic size to a regime where increasing the heterogeneity in risk tolerance results in an increase in epidemic size. This result also does not have to necessarily be due to a dramatic shift in parameters. From Figure \ref{fig:heterogeneity}, we see this shift can arise from solely varying the fraction of the population with low risk tolerance by a small amount.

The intuition underlying this result is that when the average adoption rate ($\lambda_{average}$), which is expressed as a (geometric or arithmetic) weighted mean of the adoption rates of the two groups, is fixed at a constant level, then the epidemic size can be suppressed through either varying the fraction of the population in each group or through varying each group’s adoption rate. When risk-averse people make up a smaller fraction of the population than risk-taking people, then it would be more beneficial in reducing epidemic size to have the adoption rates of the two groups be more similar (i.e. more homogeneous) as that would imply risk-taking people (which are then the majority of the population) would have a similar adoption rate to risk-averse people. As an increasingly larger fraction of the population is composed of risk-averse people, then it becomes increasingly beneficial in reducing epidemic size to have the adoption rates of the two groups be more different (i.e. more heterogeneous) as the deleterious effects of highly risk-taking people (which are then the minority of the population) can be mitigated by the large presence of risk-averse people.

\subsubsection*{Epidemic Size Does Not Necessarily Decrease with an Increasing Number of Risk-Averse People}

General intuition would suggest that as one increases the number of risk-averse people in the population that the overall epidemic size would go down. However, the introduction of the adaptive behavior mechanism allows for regimes where this is no longer strictly the case. Thus, it is no longer a guarantee that decreasing the population's overall risk tolerance will always improve epidemic outcomes.

\begin{figure*}[t!]
    \centering
    \begin{subfigure}[t]{0.5\textwidth}
        \centering
        \includegraphics[scale=.2]{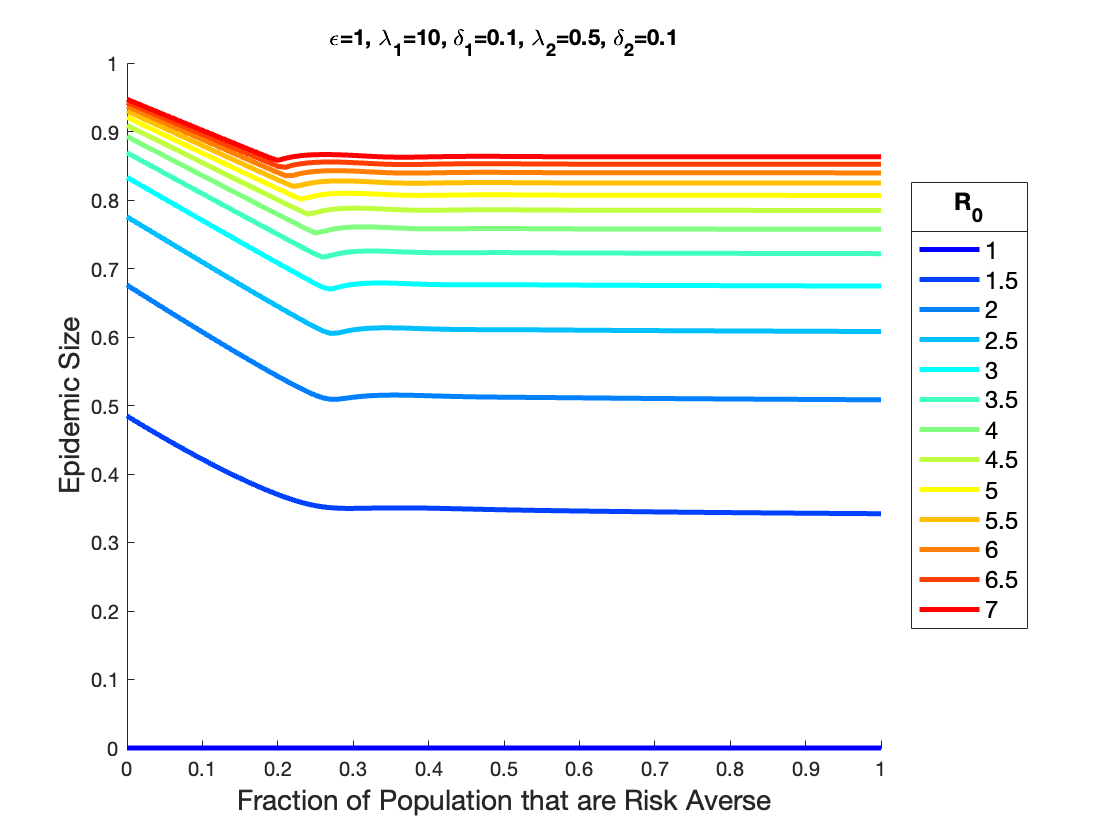}
        \caption{} \label{fig:e1}
    \end{subfigure}%
    ~ 
    \begin{subfigure}[t]{0.5\textwidth}
        \centering
        \includegraphics[scale=.2]{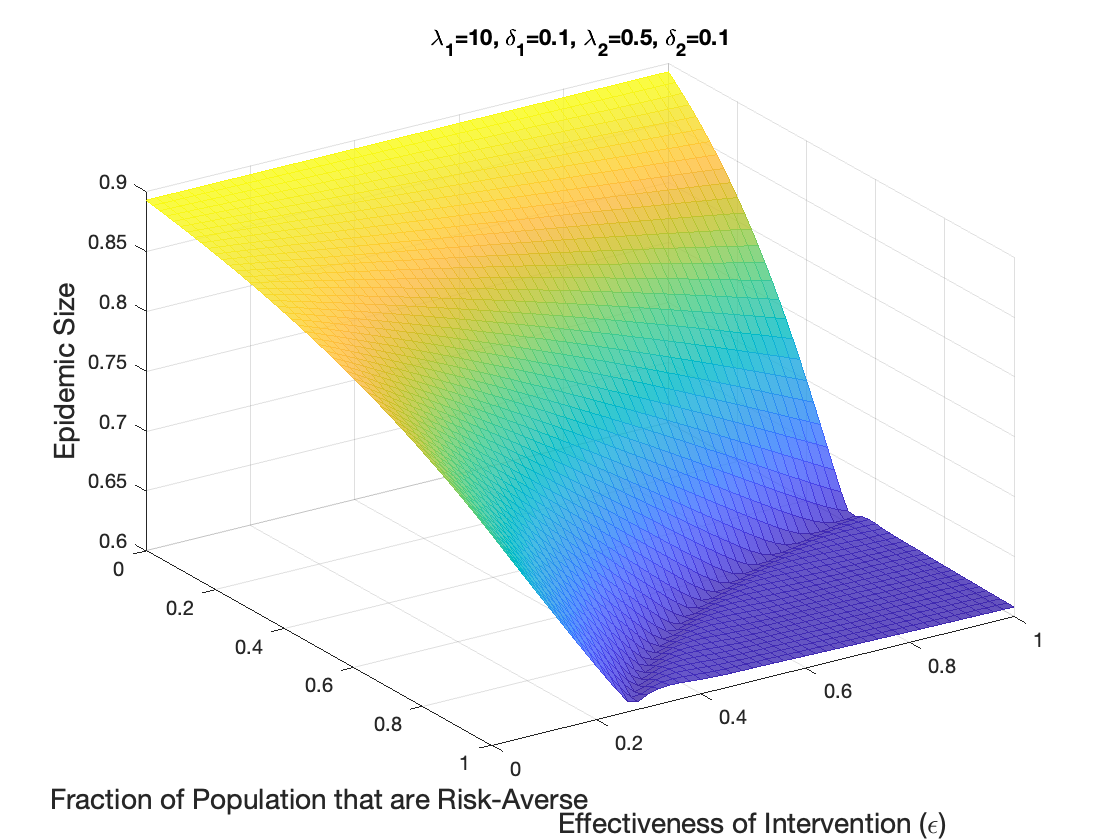}
        \caption{}
    \end{subfigure}
    \caption{(a) Final epidemic size as a function of the proportion of the population that are risk averse ($S_1$). Model parameters and initial conditions for the simulation are $\epsilon = 1, \lambda_1 = 10, \lambda_2 = 0.5, \delta_1 = 0.1, \delta_2 = 0.1, I(0) = 10^{-7}, x_{S_2}(0)=1-x_{S_1}-I(0), P_1(0)=P_2(0)=0, R(0)=0$. (b) Same as in (a) except now the effectiveness of the intervention ($\epsilon$) is allowed to vary.} \label{fig:e2}
\end{figure*}

Consider Figure \ref{fig:e1}, where we find a small region after the transition to the underdamped regime where there is an increase in the epidemic size when increasing the fraction of the population that are risk-averse.

If we also vary the effectiveness of the intervention as an additional axis (Figure \ref{fig:e2}b), we observe that there is a small trench in the threshold region surrounding the plateau area. This double descent suggests that the landscape can potentially be quite complicated when risk tolerance in the population is partitioned into even more groups.

\section*{Discussion and Conclusions}
In this paper, we have proposed a simple model to model heterogeneity in risk tolerance levels in the population. We find that including a behavioral mechanism for adopting interventions that adapts with the level of infections greatly expands the variety in epidemic dynamics and outcomes that can occur. 

The general picture from the findings suggest that epidemic dynamics under adaptive intervention adoption fall into either an underdamped regime or an overdamped regime. The underdamped regime has a special property in which the epidemic size equals the herd immunity threshold exactly, which means no epidemic overshoot occurs. The system can be driven into this regime when protection mechanisms (such as numbers of risk-averse people, intervention effectiveness, and duration of intervention usage) are increased to a sufficiently high level. This regime is also marked by damped oscillations in the phase space of infecteds and susceptibles. In direct contrast, the overdamped regime closely resembles the dynamics of a simple $SIR$ model without behavior, in which there are no oscillations and a non-zero overshoot, which makes the epidemic size greater than the herd immunity threshold. 

\begin{figure}[ht]
\centering
\includegraphics[scale=0.35]{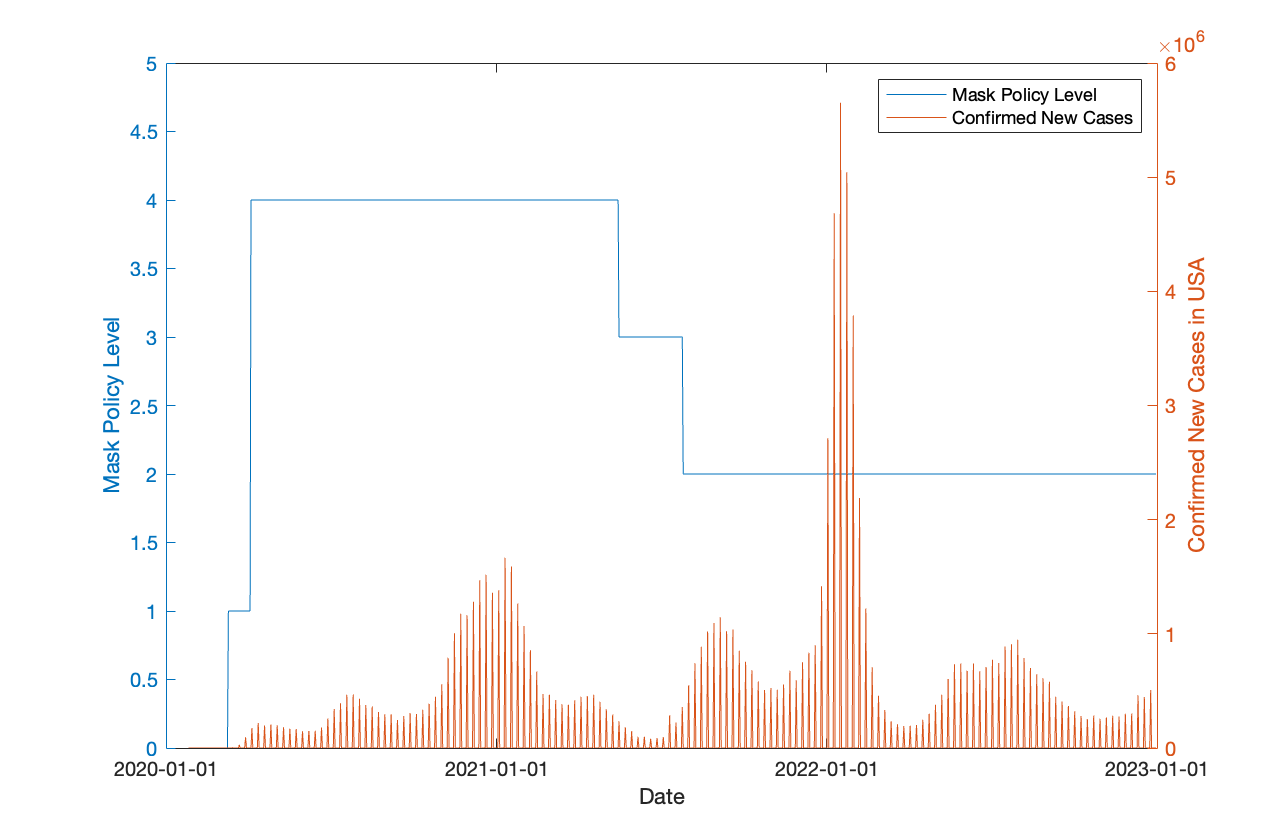}
\caption{Time series data for masking policy and confirmed new cases in the USA for the first two years of the COVID pandemic. Data extracted from \cite{mathieu_coronavirus_2020, hale_global_2021}.} \label{fig:maskPolicy}
\end{figure}

We have looked for some evidence in the historical data on outbreaks for these damped oscillations due to cycles in adoption and relaxation of interventions. While such data is in very limited supply, previous analysis suggested that relaxation of social distancing measures may have led to multiple waves of infection in the Spanish flu of the early 20th century \cite{hatchett_public_2007, caley_quantifying_2007}. Dating back to the time of the bubonic plague, there is data from an outbreak in 1636 in the parish of St. Martin in the Fields, which showed how relaxation of quarantining and isolation measures lead to a smaller secondary wave of infections \cite{newman_shutt_2012}. We also see some evidence from time-series data from early in the COVID pandemic on masking policy \cite{hale_global_2021}. Taking mask policy as a proxy for the general level of mask usage \cite{hale_global_2021}, we can observe the policy level in the United States as it relates to the overall incidence in COVID infections during that same period \cite{mathieu_coronavirus_2020}, which we take as a rough proxy for the number of infected people. We see that a relaxation in the mask policy from its strictest level coincided with the rise of the Omicron variant soon afterward (Figure \ref{fig:maskPolicy}). While there are many confounding variables at play here including viral evolution that make it difficult to disentangle the contribution of individual factors, the timing suggests that behavior plays a key role in shaping the epidemic landscape as well.

The results here are reminiscent of feedback control systems commonly studied in control theory. Here the set point is the herd immunity threshold, which is determined by the basic reproduction number ($R_0$). The ability for the population to reach this set point for epidemic size without additional overshoot depends on the effectiveness of the feedback mechanism from coupling intervention usage to the number of infected people. In the model presented, the adoption of interventions is a continuous process, in which the different groups are constantly reacting to the level of infections without requiring any notion of time or thresholds. In contrast, existing research on mitigation have considered more active control where activation and intervention timing play a key role \cite{bussell_applying_2019, lauro_optimal_2021, morris_optimal_2021}. Future work may explore how to synergistically utilize both active and continuous mechanisms for control.

The inclusion of heterogeneity in risk tolerance and adaptive adoption of interventions leads to several unexpected conclusions. We find that increasing heterogeneity in risk tolerance levels in the population can lead to either an increase or decrease in the epidemic size. The direction of the trend depends nonlinearly on the composition of the population in terms of the ratio of risk-averse to risk-taking individuals and their respective intervention adoption rates. This adds to a small literature that demonstrates how heterogeneity can actually lead to a larger epidemic \cite{volz_effects_2011, espinoza_heterogeneous_2022}. 

Interestingly, these results on heterogeneity also can be used to address the question of whether distributed or centralized control of mitigation results in a smaller outbreak. Control of factors such as mobility may be more practically achieved in a more centralized and unified fashion \cite{bonaccorsi_economic_2020, espinoza_mobility_2020}, whereas a distributed approach may be more appropriate when considering factors such as speed and individual agency. In a centralized scenario, a single entity controls the dynamics. That situation has an exact correspondence to the homogeneous population considered here, where all individuals respond in unison. Distributed control allows for more localized control, such as individuals or small groups deciding if they want to mask or social distance. This corresponds to the heterogeneous scenarios considered here, where there are multiple groups each with differing risk tolerance level. The results suggest that in some scenarios a single, coordinated response would be better for mitigation, whereas in other parameter regimes, a more decentralized strategy would be more optimal. 

We also find that increasing overall protection mechanisms does not always result in a monotonic decrease in epidemic size. In scenarios when the adoption rate begins to approach the transmission rate, near the critically damped boundary a nonmonotonicity can arise. This suggests that when intervention usage and effectiveness are tenuous, the dynamics become more complex and predicting what epidemic outcomes will result becomes significantly more difficult. Understanding how these nonlinear effects combine with other biological and behavioral heterogeneities will be important to explore in future work.

\printbibliography

@article{bansal_when_2007,
	title = {When individual behaviour matters: homogeneous and network models in epidemiology},
	volume = {4},
	url = {https://royalsocietypublishing.org/doi/full/10.1098/rsif.2007.1100},
	doi = {10.1098/rsif.2007.1100},
	shorttitle = {When individual behaviour matters},
	pages = {879--891},
	number = {16},
	journaltitle = {Journal of The Royal Society Interface},
	author = {Bansal, Shweta and Grenfell, Bryan T and Meyers, Lauren Ancel},
	urldate = {2024-03-19},
	date = {2007-07-19},
	note = {Publisher: Royal Society},
}

@article{wagner_economic_2020,
	title = {Economic and Behavioral Influencers of Vaccination and Antimicrobial Use},
	volume = {8},
	issn = {2296-2565},
	url = {https://www.frontiersin.org/journals/public-health/articles/10.3389/fpubh.2020.614113/full},
	doi = {10.3389/fpubh.2020.614113},
	journaltitle = {Frontiers in Public Health},
	shortjournal = {Front. Public Health},
	author = {Wagner, Caroline E. and Prentice, Joseph A. and Saad-Roy, Chadi M. and Yang, Luojun and Grenfell, Bryan T. and Levin, Simon A. and Laxminarayan, Ramanan},
	urldate = {2024-04-19},
	date = {2020-12-21},
	note = {Publisher: Frontiers},
}

@article{wagner_modelling_2022,
	title = {Modelling vaccination strategies for {COVID}-19},
	volume = {22},
	rights = {2022 Springer Nature Limited},
	issn = {1474-1741},
	url = {https://www.nature.com/articles/s41577-022-00687-3},
	doi = {10.1038/s41577-022-00687-3},
	pages = {139--141},
	number = {3},
	journaltitle = {Nature Reviews Immunology},
	shortjournal = {Nat Rev Immunol},
	author = {Wagner, Caroline E. and Saad-Roy, Chadi M. and Grenfell, Bryan T.},
	urldate = {2024-04-19},
	date = {2022-03},
	langid = {english},
	note = {Publisher: Nature Publishing Group},
}

@article{traulsen_individual_2023,
	title = {Individual costs and societal benefits of interventions during the {COVID}-19 pandemic},
	volume = {120},
	url = {https://www.pnas.org/doi/abs/10.1073/pnas.2303546120},
	doi = {10.1073/pnas.2303546120},
	pages = {e2303546120},
	number = {24},
	journaltitle = {Proceedings of the National Academy of Sciences},
	author = {Traulsen, Arne and Levin, Simon A. and Saad-Roy, Chadi M.},
	urldate = {2024-04-19},
	date = {2023-06-13},
	note = {Publisher: Proceedings of the National Academy of Sciences},
}

@article{espinoza_heterogeneous_2022,
	title = {Heterogeneous adaptive behavioral responses may increase epidemic burden},
	volume = {12},
	rights = {2022 The Author(s)},
	issn = {2045-2322},
	url = {https://www.nature.com/articles/s41598-022-15444-8},
	doi = {10.1038/s41598-022-15444-8},
	pages = {11276},
	number = {1},
	journaltitle = {Scientific Reports},
	shortjournal = {Sci Rep},
	author = {Espinoza, Baltazar and Swarup, Samarth and Barrett, Christopher L. and Marathe, Madhav},
	urldate = {2024-04-19},
	date = {2022-07-04},
	langid = {english},
	note = {Publisher: Nature Publishing Group},
}

@article{qiu_understanding_2022,
	title = {Understanding the coevolution of mask wearing and epidemics: A network perspective},
	volume = {119},
	url = {https://www.pnas.org/doi/abs/10.1073/pnas.2123355119},
	doi = {10.1073/pnas.2123355119},
	shorttitle = {Understanding the coevolution of mask wearing and epidemics},
	pages = {e2123355119},
	number = {26},
	journaltitle = {Proceedings of the National Academy of Sciences},
	author = {Qiu, Zirou and Espinoza, Baltazar and Vasconcelos, Vitor V. and Chen, Chen and Constantino, Sara M. and Crabtree, Stefani A. and Yang, Luojun and Vullikanti, Anil and Chen, Jiangzhuo and Weibull, Jörgen and Basu, Kaushik and Dixit, Avinash and Levin, Simon A. and Marathe, Madhav V.},
	urldate = {2024-04-19},
	date = {2022-06-28},
	note = {Publisher: Proceedings of the National Academy of Sciences},
}

@article{saad-roy_dynamics_2023,
	title = {Dynamics in a behavioral–epidemiological model for individual adherence to a nonpharmaceutical intervention},
	volume = {120},
	url = {https://www.pnas.org/doi/abs/10.1073/pnas.2311584120},
	doi = {10.1073/pnas.2311584120},
	pages = {e2311584120},
	number = {44},
	journaltitle = {Proceedings of the National Academy of Sciences},
	author = {Saad-Roy, Chadi M. and Traulsen, Arne},
	urldate = {2024-04-19},
	date = {2023-10-31},
	note = {Publisher: Proceedings of the National Academy of Sciences},
}

@article{yang_sociocultural_2022,
	title = {Sociocultural determinants of global mask-wearing behavior},
	volume = {119},
	rights = {Copyright © 2022 the Author(s). Published by {PNAS}.},
	url = {https://www.pnas.org/doi/abs/10.1073/pnas.2213525119},
	doi = {10.1073/pnas.2213525119},
	pages = {e2213525119},
	number = {41},
	journaltitle = {Proceedings of the National Academy of Sciences},
	author = {Yang, Luojun and Constantino, Sara M. and Grenfell, Bryan T. and Weber, Elke U. and Levin, Simon A. and Vasconcelos, Vítor V.},
	urldate = {2024-04-19},
	date = {2022-10-11},
	note = {Company: National Academy of Sciences
Distributor: National Academy of Sciences
Institution: National Academy of Sciences
Label: National Academy of Sciences
Publisher: Proceedings of the National Academy of Sciences},
}

@article{smith_covid-19_2023,
	title = {{COVID}-19 Mitigation Among College Students: Social Influences, Behavioral Spillover, and Antibody Results},
	volume = {38},
	issn = {1041-0236},
	url = {https://doi.org/10.1080/10410236.2022.2049047},
	doi = {10.1080/10410236.2022.2049047},
	shorttitle = {{COVID}-19 Mitigation Among College Students},
	pages = {2002--2011},
	number = {10},
	journaltitle = {Health Communication},
	author = {Smith, Rachel A. and Small, Meg L. and Bharti, Nita and {DeMatte}, Samuel J. and Lennon, Robert P. and Ferrari, Matthew J.},
	urldate = {2024-04-19},
	date = {2023-08-24},
	pmid = {35317696},
	note = {Publisher: Routledge
\_eprint: https://doi.org/10.1080/10410236.2022.2049047},
}

@article{bergstrom_human_2024,
	title = {Human behavior and disease dynamics},
	volume = {121},
	url = {https://www.pnas.org/doi/abs/10.1073/pnas.2317211120},
	doi = {10.1073/pnas.2317211120},
	pages = {e2317211120},
	number = {1},
	journaltitle = {Proceedings of the National Academy of Sciences},
	author = {Bergstrom, Carl T. and Hanage, William P.},
	urldate = {2024-04-19},
	date = {2024-01-02},
	note = {Publisher: Proceedings of the National Academy of Sciences},
}

@article{espinoza_asymptomatic_2021,
	title = {Asymptomatic individuals can increase the final epidemic size under adaptive human behavior},
	volume = {11},
	rights = {2021 The Author(s)},
	issn = {2045-2322},
	url = {https://www.nature.com/articles/s41598-021-98999-2},
	doi = {10.1038/s41598-021-98999-2},
	pages = {19744},
	number = {1},
	journaltitle = {Scientific Reports},
	shortjournal = {Sci Rep},
	author = {Espinoza, Baltazar and Marathe, Madhav and Swarup, Samarth and Thakur, Mugdha},
	urldate = {2024-04-19},
	date = {2021-10-05},
	langid = {english},
	note = {Publisher: Nature Publishing Group},
}

@incollection{murray_chapter_2016,
	title = {Chapter Two - The Behavioral Immune System: Implications for Social Cognition, Social Interaction, and Social Influence},
	volume = {53},
	url = {https://www.sciencedirect.com/science/article/pii/S0065260115000246},
	shorttitle = {Chapter Two - The Behavioral Immune System},
	pages = {75--129},
	booktitle = {Advances in Experimental Social Psychology},
	publisher = {Academic Press},
	author = {Murray, Damian R. and Schaller, Mark},
	editor = {Olson, James M. and Zanna, Mark P.},
	urldate = {2024-04-19},
	date = {2016-01-01},
	doi = {10.1016/bs.aesp.2015.09.002},
}

@article{funk_modelling_2010,
	title = {Modelling the influence of human behaviour on the spread of infectious diseases: a review},
	volume = {7},
	url = {https://royalsocietypublishing.org/doi/10.1098/rsif.2010.0142},
	doi = {10.1098/rsif.2010.0142},
	shorttitle = {Modelling the influence of human behaviour on the spread of infectious diseases},
	pages = {1247--1256},
	number = {50},
	journaltitle = {Journal of The Royal Society Interface},
	author = {Funk, Sebastian and Salathé, Marcel and Jansen, Vincent A. A.},
	urldate = {2024-04-19},
	date = {2010-05-26},
	note = {Publisher: Royal Society},
}

@article{brzezinski_science_2021,
	title = {Science skepticism reduced compliance with {COVID}-19 shelter-in-place policies in the United States},
	volume = {5},
	rights = {2021 The Author(s), under exclusive licence to Springer Nature Limited},
	issn = {2397-3374},
	url = {https://www.nature.com/articles/s41562-021-01227-0},
	doi = {10.1038/s41562-021-01227-0},
	pages = {1519--1527},
	number = {11},
	journaltitle = {Nature Human Behaviour},
	shortjournal = {Nat Hum Behav},
	author = {Brzezinski, Adam and Kecht, Valentin and Van Dijcke, David and Wright, Austin L.},
	urldate = {2024-04-19},
	date = {2021-11},
	langid = {english},
	note = {Publisher: Nature Publishing Group},
}

@article{kleitman_comply_2021,
	title = {To comply or not comply? A latent profile analysis of behaviours and attitudes during the {COVID}-19 pandemic},
	volume = {16},
	issn = {1932-6203},
	url = {https://journals.plos.org/plosone/article?id=10.1371/journal.pone.0255268},
	doi = {10.1371/journal.pone.0255268},
	shorttitle = {To comply or not comply?},
	pages = {e0255268},
	number = {7},
	journaltitle = {{PLOS} {ONE}},
	shortjournal = {{PLOS} {ONE}},
	author = {Kleitman, Sabina and Fullerton, Dayna J. and Zhang, Lisa M. and Blanchard, Matthew D. and Lee, Jihyun and Stankov, Lazar and Thompson, Valerie},
	urldate = {2024-04-19},
	date = {2021-07-29},
	langid = {english},
	note = {Publisher: Public Library of Science},
}

@article{fenichel_adaptive_2011,
	title = {Adaptive human behavior in epidemiological models},
	volume = {108},
	url = {https://www.pnas.org/doi/full/10.1073/pnas.1011250108},
	doi = {10.1073/pnas.1011250108},
	pages = {6306--6311},
	number = {15},
	journaltitle = {Proceedings of the National Academy of Sciences},
	author = {Fenichel, Eli P. and Castillo-Chavez, Carlos and Ceddia, M. G. and Chowell, Gerardo and Parra, Paula A. Gonzalez and Hickling, Graham J. and Holloway, Garth and Horan, Richard and Morin, Benjamin and Perrings, Charles and Springborn, Michael and Velazquez, Leticia and Villalobos, Cristina},
	urldate = {2024-04-19},
	date = {2011-04-12},
	note = {Publisher: Proceedings of the National Academy of Sciences},
}

@article{betsch_social_2020,
	title = {Social and behavioral consequences of mask policies during the {COVID}-19 pandemic},
	volume = {117},
	url = {https://www.pnas.org/doi/full/10.1073/pnas.2011674117},
	doi = {10.1073/pnas.2011674117},
	pages = {21851--21853},
	number = {36},
	journaltitle = {Proceedings of the National Academy of Sciences},
	author = {Betsch, Cornelia and Korn, Lars and Sprengholz, Philipp and Felgendreff, Lisa and Eitze, Sarah and Schmid, Philipp and Böhm, Robert},
	urldate = {2024-04-19},
	date = {2020-09-08},
	note = {Publisher: Proceedings of the National Academy of Sciences},
}

@article{wong_paradox_2020,
	title = {The paradox of trust: perceived risk and public compliance during the {COVID}-19 pandemic in Singapore},
	volume = {23},
	issn = {1366-9877},
	url = {https://doi.org/10.1080/13669877.2020.1756386},
	doi = {10.1080/13669877.2020.1756386},
	shorttitle = {The paradox of trust},
	pages = {1021--1030},
	number = {7},
	journaltitle = {Journal of Risk Research},
	author = {Wong, Catherine Mei Ling and Jensen, Olivia},
	urldate = {2024-04-19},
	date = {2020-08-02},
	note = {Publisher: Routledge
\_eprint: https://doi.org/10.1080/13669877.2020.1756386},
}

@article{machingaidze_understanding_2021,
	title = {Understanding {COVID}-19 vaccine hesitancy},
	volume = {27},
	rights = {2021 Springer Nature America, Inc.},
	issn = {1546-170X},
	url = {https://www.nature.com/articles/s41591-021-01459-7},
	doi = {10.1038/s41591-021-01459-7},
	pages = {1338--1339},
	number = {8},
	journaltitle = {Nature Medicine},
	shortjournal = {Nat Med},
	author = {Machingaidze, Shingai and Wiysonge, Charles Shey},
	urldate = {2024-04-19},
	date = {2021-08},
	langid = {english},
	note = {Publisher: Nature Publishing Group},
}

@article{kramer_infection_2021,
	title = {Infection threat shapes our social instincts},
	volume = {75},
	issn = {1432-0762},
	url = {https://doi.org/10.1007/s00265-021-02975-9},
	doi = {10.1007/s00265-021-02975-9},
	pages = {47},
	number = {3},
	journaltitle = {Behavioral Ecology and Sociobiology},
	shortjournal = {Behav Ecol Sociobiol},
	author = {Kramer, Peter and Bressan, Paola},
	urldate = {2024-04-19},
	date = {2021-02-10},
	langid = {english},
}

@article{fedele_covid-19_2021,
	title = {{COVID}-19 vaccine hesitancy: a survey in a population highly compliant to common vaccinations},
	volume = {17},
	issn = {2164-5515},
	url = {https://doi.org/10.1080/21645515.2021.1928460},
	doi = {10.1080/21645515.2021.1928460},
	shorttitle = {{COVID}-19 vaccine hesitancy},
	pages = {3348--3354},
	number = {10},
	journaltitle = {Human Vaccines \& Immunotherapeutics},
	author = {Fedele, Flora and Aria, Massimo and Esposito, Valeria and Micillo, Maria and Cecere, Gaetano and Spano, Maria and De Marco, Giulio},
	urldate = {2024-04-19},
	date = {2021-10-03},
	pmid = {34096836},
	note = {Publisher: Taylor \& Francis
\_eprint: https://doi.org/10.1080/21645515.2021.1928460},
}

@article{maxmen_covid_2021,
	title = {The {COVID} lab-leak hypothesis: what scientists do and don’t know},
	volume = {594},
	rights = {2021 Nature},
	url = {https://www.nature.com/articles/d41586-021-01529-3},
	doi = {10.1038/d41586-021-01529-3},
	shorttitle = {The {COVID} lab-leak hypothesis},
	pages = {313--315},
	number = {7863},
	journaltitle = {Nature},
	author = {Maxmen, Amy and Mallapaty, Smriti},
	urldate = {2024-04-19},
	date = {2021-06-08},
	langid = {english},
	note = {Bandiera\_abtest: a
Cg\_type: News Explainer
Publisher: Nature Publishing Group
Subject\_term: Public health, Epidemiology, Diseases, Politics, {SARS}-{CoV}-2},
}

@article{morens_concept_2022,
	title = {The Concept of Classical Herd Immunity May Not Apply to {COVID}-19},
	volume = {226},
	issn = {0022-1899},
	url = {https://doi.org/10.1093/infdis/jiac109},
	doi = {10.1093/infdis/jiac109},
	pages = {195--198},
	number = {2},
	journaltitle = {The Journal of Infectious Diseases},
	shortjournal = {The Journal of Infectious Diseases},
	author = {Morens, David M and Folkers, Gregory K and Fauci, Anthony S},
	urldate = {2024-04-19},
	date = {2022-07-15},
}

@incollection{osterholm_preparing_2020,
	title = {Preparing for the Next Pandemic},
	isbn = {978-1-00-314140-2},
	booktitle = {The Covid-19 Reader},
	publisher = {Routledge},
	author = {Osterholm, Michael T.},
	date = {2020},
	note = {Num Pages: 10},
}

@article{morse_prediction_2012,
	title = {Prediction and prevention of the next pandemic zoonosis},
	volume = {380},
	issn = {0140-6736, 1474-547X},
	url = {https://www.thelancet.com/journals/lancet/article/PIIS0140-6736(12)61684-5/fulltext},
	doi = {10.1016/S0140-6736(12)61684-5},
	pages = {1956--1965},
	number = {9857},
	journaltitle = {The Lancet},
	shortjournal = {The Lancet},
	author = {Morse, Stephen S. and Mazet, Jonna {AK} and Woolhouse, Mark and Parrish, Colin R. and Carroll, Dennis and Karesh, William B. and Zambrana-Torrelio, Carlos and Lipkin, W. Ian and Daszak, Peter},
	urldate = {2024-04-19},
	date = {2012-12-01},
	pmid = {23200504},
	note = {Publisher: Elsevier},
}

@article{franzen_fatigue_2021,
	title = {Fatigue during the {COVID}-19 pandemic: Evidence of social distancing adherence from a panel study of young adults in Switzerland},
	volume = {16},
	issn = {1932-6203},
	url = {https://journals.plos.org/plosone/article?id=10.1371/journal.pone.0261276},
	doi = {10.1371/journal.pone.0261276},
	shorttitle = {Fatigue during the {COVID}-19 pandemic},
	pages = {e0261276},
	number = {12},
	journaltitle = {{PLOS} {ONE}},
	shortjournal = {{PLOS} {ONE}},
	author = {Franzen, Axel and Wöhner, Fabienne},
	urldate = {2024-04-19},
	date = {2021-12-10},
	langid = {english},
	note = {Publisher: Public Library of Science},
}

@article{jorgensen_pandemic_2022,
	title = {Pandemic fatigue fueled political discontent during the {COVID}-19 pandemic},
	volume = {119},
	url = {https://www.pnas.org/doi/abs/10.1073/pnas.2201266119},
	doi = {10.1073/pnas.2201266119},
	pages = {e2201266119},
	number = {48},
	journaltitle = {Proceedings of the National Academy of Sciences},
	author = {Jørgensen, Frederik and Bor, Alexander and Rasmussen, Magnus Storm and Lindholt, Marie Fly and Petersen, Michael Bang},
	urldate = {2024-04-19},
	date = {2022-11-29},
	note = {Publisher: Proceedings of the National Academy of Sciences},
}

@article{cheok_appropriate_2021,
	title = {Appropriate attitude promotes mask wearing in spite of a significant experience of varying discomfort},
	volume = {26},
	issn = {2468-0451},
	url = {https://www.sciencedirect.com/science/article/pii/S246804512100002X},
	doi = {10.1016/j.idh.2021.01.002},
	pages = {145--151},
	number = {2},
	journaltitle = {Infection, Disease \& Health},
	shortjournal = {Infection, Disease \& Health},
	author = {Cheok, Gideon J. W. and Gatot, Cheryl and Sim, Craigven H. S. and Ng, Y. H. and Tay, Kenny X. K. and Howe, T. S. and Koh, Joyce S. B.},
	urldate = {2024-04-19},
	date = {2021-05-01},
}

@article{edmunds_evaluating_1999,
	title = {Evaluating the cost-effectiveness of vaccination programmes: a dynamic perspective},
	volume = {18},
	issn = {1097-0258},
	url = {https://onlinelibrary.wiley.com/doi/abs/10.1002/%28SICI%291097-0258%2819991215%2918%3A23%3C3263%3A%3AAID-SIM315%3E3.0.CO%3B2-3},
	doi = {10.1002/(SICI)1097-0258(19991215)18:23<3263::AID-SIM315>3.0.CO;2-3},
	shorttitle = {Evaluating the cost-effectiveness of vaccination programmes},
	pages = {3263--3282},
	number = {23},
	journaltitle = {Statistics in Medicine},
	author = {Edmunds, W. J. and Medley, G. F. and Nokes, D. J.},
	urldate = {2024-04-29},
	date = {1999},
	langid = {english},
	note = {\_eprint: https://onlinelibrary.wiley.com/doi/pdf/10.1002/\%28SICI\%291097-0258\%2819991215\%2918\%3A23\%3C3263\%3A\%3AAID-{SIM}315\%3E3.0.{CO}\%3B2-3},
}

@article{fischer_mask_2021,
	title = {Mask adherence and rate of {COVID}-19 across the United States},
	volume = {16},
	issn = {1932-6203},
	url = {https://journals.plos.org/plosone/article?id=10.1371/journal.pone.0249891},
	doi = {10.1371/journal.pone.0249891},
	pages = {e0249891},
	number = {4},
	journaltitle = {{PLOS} {ONE}},
	shortjournal = {{PLOS} {ONE}},
	author = {Fischer, Charlie B. and Adrien, Nedghie and Silguero, Jeremiah J. and Hopper, Julianne J. and Chowdhury, Abir I. and Werler, Martha M.},
	urldate = {2024-04-29},
	date = {2021-04-14},
	langid = {english},
	note = {Publisher: Public Library of Science},
}

@article{miller_epidemic_2007,
	title = {Epidemic size and probability in populations with heterogeneous infectivity and susceptibility},
	volume = {76},
	url = {https://link.aps.org/doi/10.1103/PhysRevE.76.010101},
	doi = {10.1103/PhysRevE.76.010101},
	pages = {010101},
	number = {1},
	journaltitle = {Physical Review E},
	shortjournal = {Phys. Rev. E},
	author = {Miller, Joel C.},
	urldate = {2024-05-03},
	date = {2007-07-10},
	note = {Publisher: American Physical Society},
}

@article{gomes_individual_2022,
	title = {Individual variation in susceptibility or exposure to {SARS}-{CoV}-2 lowers the herd immunity threshold},
	volume = {540},
	issn = {0022-5193},
	url = {https://www.sciencedirect.com/science/article/pii/S0022519322000613},
	doi = {10.1016/j.jtbi.2022.111063},
	pages = {111063},
	journaltitle = {Journal of Theoretical Biology},
	shortjournal = {Journal of Theoretical Biology},
	author = {Gomes, M. Gabriela M. and Ferreira, Marcelo U. and Corder, Rodrigo M. and King, Jessica G. and Souto-Maior, Caetano and Penha-Gonçalves, Carlos and Gonçalves, Guilherme and Chikina, Maria and Pegden, Wesley and Aguas, Ricardo},
	urldate = {2024-05-03},
	date = {2022-05-07},
}

@article{britton_mathematical_2020,
	title = {A mathematical model reveals the influence of population heterogeneity on herd immunity to {SARS}-{CoV}-2},
	volume = {369},
	url = {https://www.science.org/doi/full/10.1126/science.abc6810},
	doi = {10.1126/science.abc6810},
	pages = {846--849},
	number = {6505},
	journaltitle = {Science},
	author = {Britton, Tom and Ball, Frank and Trapman, Pieter},
	urldate = {2024-05-03},
	date = {2020-08-14},
	note = {Publisher: American Association for the Advancement of Science},
}

@article{volz_effects_2011,
	title = {Effects of Heterogeneous and Clustered Contact Patterns on Infectious Disease Dynamics},
	volume = {7},
	issn = {1553-7358},
	url = {https://journals.plos.org/ploscompbiol/article?id=10.1371/journal.pcbi.1002042},
	doi = {10.1371/journal.pcbi.1002042},
	pages = {e1002042},
	number = {6},
	journaltitle = {{PLOS} Computational Biology},
	shortjournal = {{PLOS} Computational Biology},
	author = {Volz, Erik M. and Miller, Joel C. and Galvani, Alison and Meyers, Lauren Ancel},
	urldate = {2024-05-03},
	date = {2011-06-02},
	langid = {english},
	note = {Publisher: Public Library of Science},
}

@article{allard_role_2023,
	title = {The Role of Directionality, Heterogeneity, and Correlations in Epidemic Risk and Spread},
	volume = {65},
	issn = {0036-1445},
	url = {https://epubs.siam.org/doi/abs/10.1137/20M1383811},
	doi = {10.1137/20M1383811},
	pages = {471--492},
	number = {2},
	journaltitle = {{SIAM} Review},
	shortjournal = {{SIAM} Rev.},
	author = {Allard, Antoine and Moore, Cristopher and Scarpino, Samuel V. and Althouse, Benjamin M. and Hébert-Dufresne, Laurent},
	urldate = {2024-05-03},
	date = {2023-05-04},
	note = {Publisher: Society for Industrial and Applied Mathematics},
}

@article{barak_experience_2022,
	title = {Experience of the {COVID}-19 pandemic in Wuhan leads to a lasting increase in social distancing},
	volume = {12},
	rights = {2022 The Author(s)},
	issn = {2045-2322},
	url = {https://www.nature.com/articles/s41598-022-23019-w},
	doi = {10.1038/s41598-022-23019-w},
	pages = {18457},
	number = {1},
	journaltitle = {Scientific Reports},
	shortjournal = {Sci Rep},
	author = {Barak, Darija and Gallo, Edoardo and Rong, Ke and Tang, Ke and Du, Wei},
	urldate = {2024-05-03},
	date = {2022-11-02},
	langid = {english},
	note = {Publisher: Nature Publishing Group},
}

@article{offer-westort_battling_2024,
	title = {Battling the coronavirus ‘infodemic’ among social media users in Kenya and Nigeria},
	rights = {2024 The Author(s), under exclusive licence to Springer Nature Limited},
	issn = {2397-3374},
	url = {https://www.nature.com/articles/s41562-023-01810-7},
	doi = {10.1038/s41562-023-01810-7},
	pages = {1--12},
	journaltitle = {Nature Human Behaviour},
	shortjournal = {Nat Hum Behav},
	author = {Offer-Westort, Molly and Rosenzweig, Leah R. and Athey, Susan},
	urldate = {2024-05-03},
	date = {2024-03-18},
	langid = {english},
	note = {Publisher: Nature Publishing Group},
}

@article{loomba_measuring_2021,
	title = {Measuring the impact of {COVID}-19 vaccine misinformation on vaccination intent in the {UK} and {USA}},
	volume = {5},
	rights = {2021 The Author(s), under exclusive licence to Springer Nature Limited},
	issn = {2397-3374},
	url = {https://www.nature.com/articles/s41562-021-01056-1},
	doi = {10.1038/s41562-021-01056-1},
	pages = {337--348},
	number = {3},
	journaltitle = {Nature Human Behaviour},
	shortjournal = {Nat Hum Behav},
	author = {Loomba, Sahil and de Figueiredo, Alexandre and Piatek, Simon J. and de Graaf, Kristen and Larson, Heidi J.},
	urldate = {2024-05-03},
	date = {2021-03},
	langid = {english},
	note = {Publisher: Nature Publishing Group},
}

@article{gallotti_assessing_2020,
	title = {Assessing the risks of ‘infodemics’ in response to {COVID}-19 epidemics},
	volume = {4},
	rights = {2020 The Author(s), under exclusive licence to Springer Nature Limited},
	issn = {2397-3374},
	url = {https://www.nature.com/articles/s41562-020-00994-6},
	doi = {10.1038/s41562-020-00994-6},
	pages = {1285--1293},
	number = {12},
	journaltitle = {Nature Human Behaviour},
	shortjournal = {Nat Hum Behav},
	author = {Gallotti, Riccardo and Valle, Francesco and Castaldo, Nicola and Sacco, Pierluigi and De Domenico, Manlio},
	urldate = {2024-05-03},
	date = {2020-12},
	langid = {english},
	note = {Publisher: Nature Publishing Group},
}

@article{salomon_us_2021,
	title = {The {US} {COVID}-19 Trends and Impact Survey: Continuous real-time measurement of {COVID}-19 symptoms, risks, protective behaviors, testing, and vaccination},
	volume = {118},
	url = {https://www.pnas.org/doi/abs/10.1073/pnas.2111454118},
	doi = {10.1073/pnas.2111454118},
	shorttitle = {The {US} {COVID}-19 Trends and Impact Survey},
	pages = {e2111454118},
	number = {51},
	journaltitle = {Proceedings of the National Academy of Sciences},
	author = {Salomon, Joshua A. and Reinhart, Alex and Bilinski, Alyssa and Chua, Eu Jing and La Motte-Kerr, Wichada and Rönn, Minttu M. and Reitsma, Marissa B. and Morris, Katherine A. and {LaRocca}, Sarah and Farag, Tamer H. and Kreuter, Frauke and Rosenfeld, Roni and Tibshirani, Ryan J.},
	urldate = {2024-05-07},
	date = {2021-12-21},
	note = {Publisher: Proceedings of the National Academy of Sciences},
}

@article{rader_mask-wearing_2021,
	title = {Mask-wearing and control of {SARS}-{CoV}-2 transmission in the {USA}: a cross-sectional study},
	volume = {3},
	issn = {2589-7500},
	url = {https://www.thelancet.com/journals/landig/article/PIIS2589-7500(20)30293-4/fulltext?ftag=MSFd61514f&fbclid=IwAR08DO9occY2DYuQT_G60V668nBRxRg3BL4hjj4-Yu0qjneqmADQjJL5CxE#%20},
	doi = {10.1016/S2589-7500(20)30293-4},
	shorttitle = {Mask-wearing and control of {SARS}-{CoV}-2 transmission in the {USA}},
	pages = {e148--e157},
	number = {3},
	journaltitle = {The Lancet Digital Health},
	shortjournal = {The Lancet Digital Health},
	author = {Rader, Benjamin and White, Laura F. and Burns, Michael R. and Chen, Jack and Brilliant, Joseph and Cohen, Jon and Shaman, Jeffrey and Brilliant, Larry and Kraemer, Moritz U. G. and Hawkins, Jared B. and Scarpino, Samuel V. and Astley, Christina M. and Brownstein, John S.},
	urldate = {2024-05-07},
	date = {2021-03-01},
	pmid = {33483277},
	note = {Publisher: Elsevier},
}

@article{mathieu_coronavirus_2020,
	title = {Coronavirus Pandemic ({COVID}-19)},
	url = {https://ourworldindata.org/covid-google-mobility-trends},
	journaltitle = {Our World in Data},
	shortjournal = {Our World in Data},
	author = {Mathieu, Edouard and Ritchie, Hannah and Rodés-Guirao, Lucas and Appel, Cameron and Giattino, Charlie and Hasell, Joe and Macdonald, Bobbie and Dattani, Saloni and Beltekian, Diana and Ortiz-Ospina, Esteban and Roser, Max},
	urldate = {2024-05-09},
	date = {2020-03-05},
}

@article{weitz_awareness-driven_2020,
	title = {Awareness-driven behavior changes can shift the shape of epidemics away from peaks and toward plateaus, shoulders, and oscillations},
	volume = {117},
	url = {https://www.pnas.org/doi/abs/10.1073/pnas.2009911117},
	doi = {10.1073/pnas.2009911117},
	pages = {32764--32771},
	number = {51},
	journaltitle = {Proceedings of the National Academy of Sciences},
	author = {Weitz, Joshua S. and Park, Sang Woo and Eksin, Ceyhun and Dushoff, Jonathan},
	urldate = {2024-05-10},
	date = {2020-12-22},
	note = {Publisher: Proceedings of the National Academy of Sciences},
}

@article{hale_global_2021,
	title = {A global panel database of pandemic policies (Oxford {COVID}-19 Government Response Tracker)},
	volume = {5},
	rights = {2021 The Author(s), under exclusive licence to Springer Nature Limited part of Springer Nature},
	issn = {2397-3374},
	url = {https://www.nature.com/articles/s41562-021-01079-8},
	doi = {10.1038/s41562-021-01079-8},
	pages = {529--538},
	number = {4},
	journaltitle = {Nature Human Behaviour},
	shortjournal = {Nat Hum Behav},
	author = {Hale, Thomas and Angrist, Noam and Goldszmidt, Rafael and Kira, Beatriz and Petherick, Anna and Phillips, Toby and Webster, Samuel and Cameron-Blake, Emily and Hallas, Laura and Majumdar, Saptarshi and Tatlow, Helen},
	urldate = {2024-05-10},
	date = {2021-04},
	langid = {english},
	note = {Publisher: Nature Publishing Group},
}

@article{bussell_applying_2019,
	title = {Applying optimal control theory to complex epidemiological models to inform real-world disease management},
	volume = {374},
	issn = {0962-8436, 1471-2970},
	url = {https://royalsocietypublishing.org/doi/10.1098/rstb.2018.0284},
	doi = {10.1098/rstb.2018.0284},
	pages = {20180284},
	number = {1776},
	journaltitle = {Philosophical Transactions of the Royal Society B: Biological Sciences},
	shortjournal = {Phil. Trans. R. Soc. B},
	author = {Bussell, E. H. and Dangerfield, C. E. and Gilligan, C. A. and Cunniffe, N. J.},
	urldate = {2024-05-21},
	date = {2019-07-08},
	langid = {english},
}

@article{lauro_optimal_2021,
	title = {Optimal timing of one-shot interventions for epidemic control},
	volume = {17},
	issn = {1553-7358},
	url = {https://journals.plos.org/ploscompbiol/article?id=10.1371/journal.pcbi.1008763},
	doi = {10.1371/journal.pcbi.1008763},
	pages = {e1008763},
	number = {3},
	journaltitle = {{PLOS} Computational Biology},
	shortjournal = {{PLOS} Computational Biology},
	author = {Lauro, Francesco Di and Kiss, István Z. and Miller, Joel C.},
	urldate = {2024-05-21},
	date = {2021-03-18},
	langid = {english},
	note = {Publisher: Public Library of Science},
}

@article{morris_optimal_2021,
	title = {Optimal, near-optimal, and robust epidemic control},
	volume = {4},
	rights = {2021 The Author(s)},
	issn = {2399-3650},
	url = {https://www.nature.com/articles/s42005-021-00570-y},
	doi = {10.1038/s42005-021-00570-y},
	pages = {1--8},
	number = {1},
	journaltitle = {Communications Physics},
	shortjournal = {Commun Phys},
	author = {Morris, Dylan H. and Rossine, Fernando W. and Plotkin, Joshua B. and Levin, Simon A.},
	urldate = {2024-05-21},
	date = {2021-04-20},
	langid = {english},
	note = {Publisher: Nature Publishing Group},
}

@article{towers_mass_2015,
	title = {Mass Media and the Contagion of Fear: The Case of Ebola in America},
	volume = {10},
	issn = {1932-6203},
	url = {https://journals.plos.org/plosone/article?id=10.1371/journal.pone.0129179},
	doi = {10.1371/journal.pone.0129179},
	shorttitle = {Mass Media and the Contagion of Fear},
	pages = {e0129179},
	number = {6},
	journaltitle = {{PLOS} {ONE}},
	shortjournal = {{PLOS} {ONE}},
	author = {Towers, Sherry and Afzal, Shehzad and Bernal, Gilbert and Bliss, Nadya and Brown, Shala and Espinoza, Baltazar and Jackson, Jasmine and Judson-Garcia, Julia and Khan, Maryam and Lin, Michael and Mamada, Robert and Moreno, Victor M. and Nazari, Fereshteh and Okuneye, Kamaldeen and Ross, Mary L. and Rodriguez, Claudia and Medlock, Jan and Ebert, David and Castillo-Chavez, Carlos},
	urldate = {2024-05-23},
	date = {2015-06-11},
	langid = {english},
	note = {Publisher: Public Library of Science},
}

@article{espinoza_mobility_2020,
	title = {Mobility restrictions for the control of epidemics: When do they work?},
	volume = {15},
	issn = {1932-6203},
	url = {https://journals.plos.org/plosone/article?id=10.1371/journal.pone.0235731},
	doi = {10.1371/journal.pone.0235731},
	shorttitle = {Mobility restrictions for the control of epidemics},
	pages = {e0235731},
	number = {7},
	journaltitle = {{PLOS} {ONE}},
	shortjournal = {{PLOS} {ONE}},
	author = {Espinoza, Baltazar and Castillo-Chavez, Carlos and Perrings, Charles},
	urldate = {2024-05-23},
	date = {2020-07-06},
	langid = {english},
	note = {Publisher: Public Library of Science},
}

@article{bonaccorsi_economic_2020,
	title = {Economic and social consequences of human mobility restrictions under {COVID}-19},
	volume = {117},
	url = {https://www.pnas.org/doi/full/10.1073/pnas.2007658117},
	doi = {10.1073/pnas.2007658117},
	pages = {15530--15535},
	number = {27},
	journaltitle = {Proceedings of the National Academy of Sciences},
	author = {Bonaccorsi, Giovanni and Pierri, Francesco and Cinelli, Matteo and Flori, Andrea and Galeazzi, Alessandro and Porcelli, Francesco and Schmidt, Ana Lucia and Valensise, Carlo Michele and Scala, Antonio and Quattrociocchi, Walter and Pammolli, Fabio},
	urldate = {2024-05-23},
	date = {2020-07-07},
	note = {Publisher: Proceedings of the National Academy of Sciences},
}

@article{lu_collectivism_2021,
	title = {Collectivism predicts mask use during {COVID}-19},
	volume = {118},
	url = {https://www.pnas.org/doi/abs/10.1073/pnas.2021793118},
	doi = {10.1073/pnas.2021793118},
	pages = {e2021793118},
	number = {23},
	journaltitle = {Proceedings of the National Academy of Sciences},
	author = {Lu, Jackson G. and Jin, Peter and English, Alexander S.},
	urldate = {2024-05-28},
	date = {2021-06-08},
	note = {Publisher: Proceedings of the National Academy of Sciences},
}

@article{caley_quantifying_2007,
	title = {Quantifying social distancing arising from pandemic influenza},
	volume = {5},
	url = {https://royalsocietypublishing.org/doi/10.1098/rsif.2007.1197},
	doi = {10.1098/rsif.2007.1197},
	pages = {631--639},
	number = {23},
	journaltitle = {Journal of The Royal Society Interface},
	author = {Caley, Peter and Philp, David J and {McCracken}, Kevin},
	urldate = {2024-06-07},
	date = {2007-10-04},
	note = {Publisher: Royal Society},
}

@article{hatchett_public_2007,
	title = {Public health interventions and epidemic intensity during the 1918 influenza pandemic},
	volume = {104},
	url = {https://www.pnas.org/doi/full/10.1073/pnas.0610941104},
	doi = {10.1073/pnas.0610941104},
	pages = {7582--7587},
	number = {18},
	journaltitle = {Proceedings of the National Academy of Sciences},
	author = {Hatchett, Richard J. and Mecher, Carter E. and Lipsitch, Marc},
	urldate = {2024-06-07},
	date = {2007-05},
	note = {Publisher: Proceedings of the National Academy of Sciences},
}

@article{newman_shutt_2012,
	title = {Shutt Up: Bubonic Plague and Quarantine in Early Modern England},
	volume = {45},
	issn = {0022-4529},
	url = {https://doi.org/10.1093/jsh/shr114},
	doi = {10.1093/jsh/shr114},
	shorttitle = {Shutt Up},
	pages = {809--834},
	number = {3},
	journaltitle = {Journal of Social History},
	shortjournal = {Journal of Social History},
	author = {Newman, Kira L. S.},
	urldate = {2024-06-08},
	date = {2012-03-01},
}

@article{tyson_timing_2020,
	title = {The Timing and Nature of Behavioural Responses Affect the Course of an Epidemic},
	volume = {82},
	issn = {1522-9602},
	url = {https://doi.org/10.1007/s11538-019-00684-z},
	doi = {10.1007/s11538-019-00684-z},
	pages = {14},
	number = {1},
	journaltitle = {Bulletin of Mathematical Biology},
	shortjournal = {Bull Math Biol},
	author = {Tyson, Rebecca C. and Hamilton, Stephanie D. and Lo, Aboubakr S. and Baumgaertner, Bert O. and Krone, Stephen M.},
	urldate = {2024-06-13},
	date = {2020-01-14},
	langid = {english},
}

\section*{Methods}
\subsection*{Defining the $\lambda$ Homogeneity Index}
The $\lambda$ homogeneity index is defined as follows. We will assume the initial condition that at the beginning of the dynamics, the total population is composed of the fraction of the population in the low risk tolerance group $x_1$, the high risk tolerance group $x_2$, or the infected class. We will assume the fraction of the population initially infected is sufficiently small so that size of the two susceptible compartments is given by $x_1$ and $1-x_1$ respectively.

We can define the average adoption rate as being either a geometric or arithmetic mean of the two adoption rates. The choice one makes is arbitrary, so we present prescriptions for both routes. In both cases, we will map the level of homogeneity to the unit interval.

\subsubsection*{Geometric Average}
Let the geometric average of the two adoption rates be given by $\lambda_{\text{Geometric Average}}$.
\begin{align}
\lambda_{\text{Geometric Average}} = \lambda_1^{x_1} \lambda_2^{1-x_1}
\end{align}
Define $\lambda_2$ as a fraction between 0 and 1 of the average $\lambda$. 
\begin{align}
\lambda_2=c\lambda_{\text{Geometric Average}}, c\in (0,1]
\end{align}
These two equations combine to define $\lambda_1$.
\begin{align}
\lambda_1=\frac{\lambda_{\text{Geometric Average}}}{c^\frac{x_1}{1-x_1}}, c\in (0,1]
\end{align}

An index value of 1 indicates $\lambda_1=\lambda_2$, while decreasing the index value towards 0 increases the difference between $\lambda_1$ and $ \lambda_2$.

\subsubsection*{Arithmetic Average}
Let the arithmetic average of the two adoption rates be given by $\lambda_{\text{Arithmetic Average}}$.
\begin{align}
\lambda_{\text{Arithmetic Average}} = x_1\lambda_1 + (1-x_1)\lambda_2
\end{align}
Define $\lambda_2$ as a fraction between 0 and 1 of the average $\lambda$. 
\begin{align}
\lambda_2=c\lambda_{\text{Arithmetic Average}}, c\in (0,1]
\end{align}
These two equations combine to define $\lambda_1$.
\begin{align}
\lambda_1=\frac{\lambda_{\text{Arithmetic Average}}(1-x_1 c)}{1-x_1}, c\in (0,1]
\end{align}

Again, an index value of 1 indicates $\lambda_1=\lambda_2$, while decreasing the index value towards 0 increases the difference between $\lambda_1$ and $ \lambda_2$.

\subsection*{Numerical Solutions and Code}

\section*{Acknowledgements}
M.M.N., A.S.F., M.A.C., and S.A.L. would like to acknowledge funding from NSF (CCF1917819, CNS-2041952, DMS-2327711), Army Research Office (W911NF-18-1-0325), and a gift from William H. Miller III. C.M.S.-R. acknowledges funding from the Miller Institute for Basic Research in Science of UC Berkeley via a Miller Research Fellowship. B.E.C. would like to acknowledge funding from NSF (IHBEM grant 2327710 and Expeditions NSF 1918656). B.T.G. would like to acknowledge the Princeton Catalysis Initiative and Princeton Precision Medicine. 

\section*{Author Contributions}
designed research, performed research, and wrote and reviewed the manuscript.

\section*{Additional information}

\newpage
\begin{center}
\textbf{\large Supplemental Material: The Complex Interplay Between Risk Tolerance and the Spread of Infectious Diseases}
\end{center}
\setcounter{figure}{0}
\renewcommand{\figurename}{Figure}
\renewcommand{\thefigure}{S\arabic{figure}}

\subsubsection*{Analysis of the Model When Individuals Respond to the Incidence Rate}
The rate at which intervention adoption occurs may be driven by individuals considering information such as the epidemic incidence rate (e.g. cases per day), the total number of infected individuals in the population (e.g. total number of active cases), and mortality rate (e.g. deaths per day) \cite{weitz_awareness-driven_2020}. Here we will consider the first case, where individuals adopt interventions based on the incidence rate for infections. Recall that the incidence rate is given by $\sum_{i=1}^n (\beta S_i I + (1-\epsilon)\beta P_i I)$. Parameterizing each person’s individual risk tolerance by $\lambda_i$, let us assume each individual adopts an intervention at rate $\lambda _i \sum_{i=1}^n (\beta S_i I + (1-\epsilon)\beta P_i I)$. Then, if there are $S_i$ number of people that behave exactly the same (i.e. have the same level of risk-aversion), then at the population scale there is a collective adoption rate of $\lambda _i S_i \sum_{i=1}^n (\beta S_i I + (1-\epsilon)\beta P_i I)$. The same reasoning holds for each of the $n$ tolerance levels. The corresponding equations for this model are given by (\ref{eqn:incmodelfirst}-\ref{eqn:incmodellast}).

\begin{align}
\frac{dS_i}{dt} &= -\beta S_i I -\lambda _i S_i \sum_{i=1}^n (\beta S_i I + (1-\epsilon)\beta P_i I) + \delta_i P_i \label{eqn:incmodelfirst}\\
\frac{dP_i}{dt} &= -(1-\epsilon)\beta P_i I - \delta_i P_i + \lambda _i S_i \sum_{i=1}^n (\beta S_i I + (1-\epsilon)\beta P_i I) \\
\frac{dI}{dt} &= - \gamma I+\sum_{i=1}^n (\beta S_i I + (1-\epsilon)\beta P_i I)\\
\frac{dR}{dt} &= \gamma I \label{eqn:incmodellast}
\end{align}

When comparing the final epidemic size and epidemic trajectories between this model and the model in the main text where individuals adopt interventions at a rate that is based on the total number of infected people (Figures \ref{fig:saturationFixedX}-\ref{fig:saturationFixedEpsilon}), the results are indistinguishable (Figures \ref{fig:incidence_saturationFixedX}-\ref{fig:incidence_saturationFixedEpsilon}).

\begin{figure}[ht]
\centering
\includegraphics[scale=0.2]{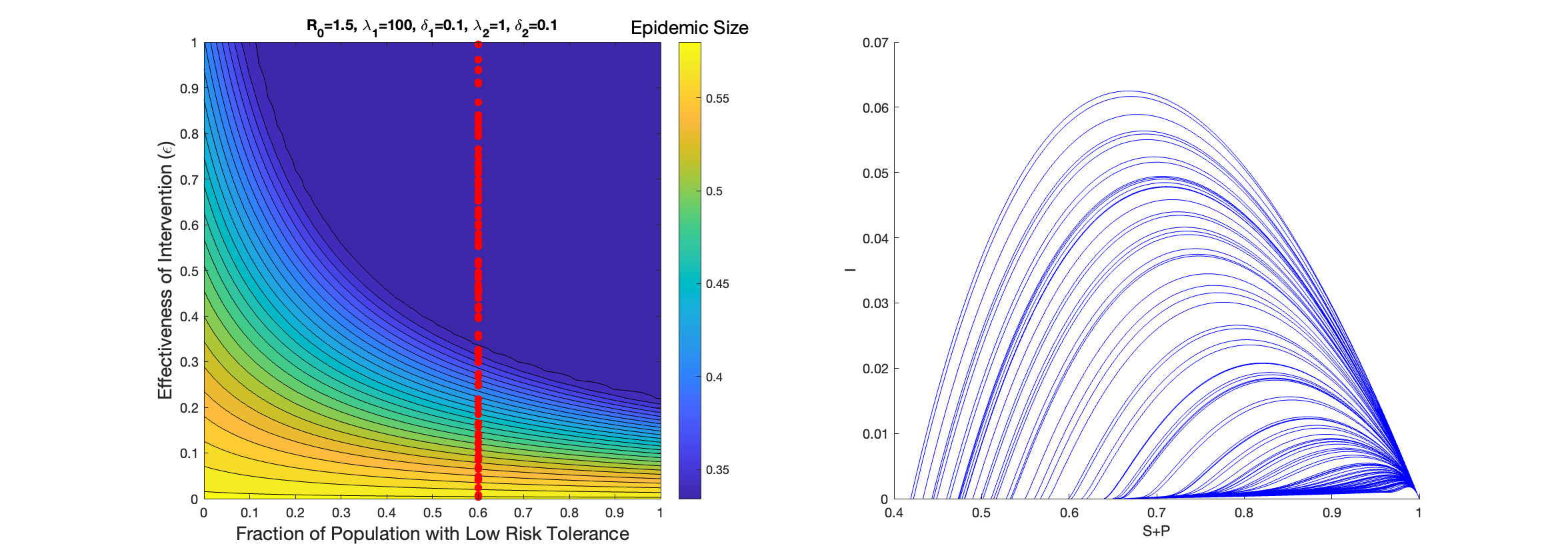}
\caption{Left. Epidemic size as a function of varying the fraction of the population that are low-risk tolerance (i.e. those with higher $\lambda$) when individuals react to incidence rate. Right. Corresponding orbits in the I versus S+P plane for the sampled points in parameter space when the fraction of the population that are low-risk tolerance has been fixed.} \label{fig:incidence_saturationFixedX}
\end{figure}

\begin{figure}[ht]
\centering
\includegraphics[scale=0.22]{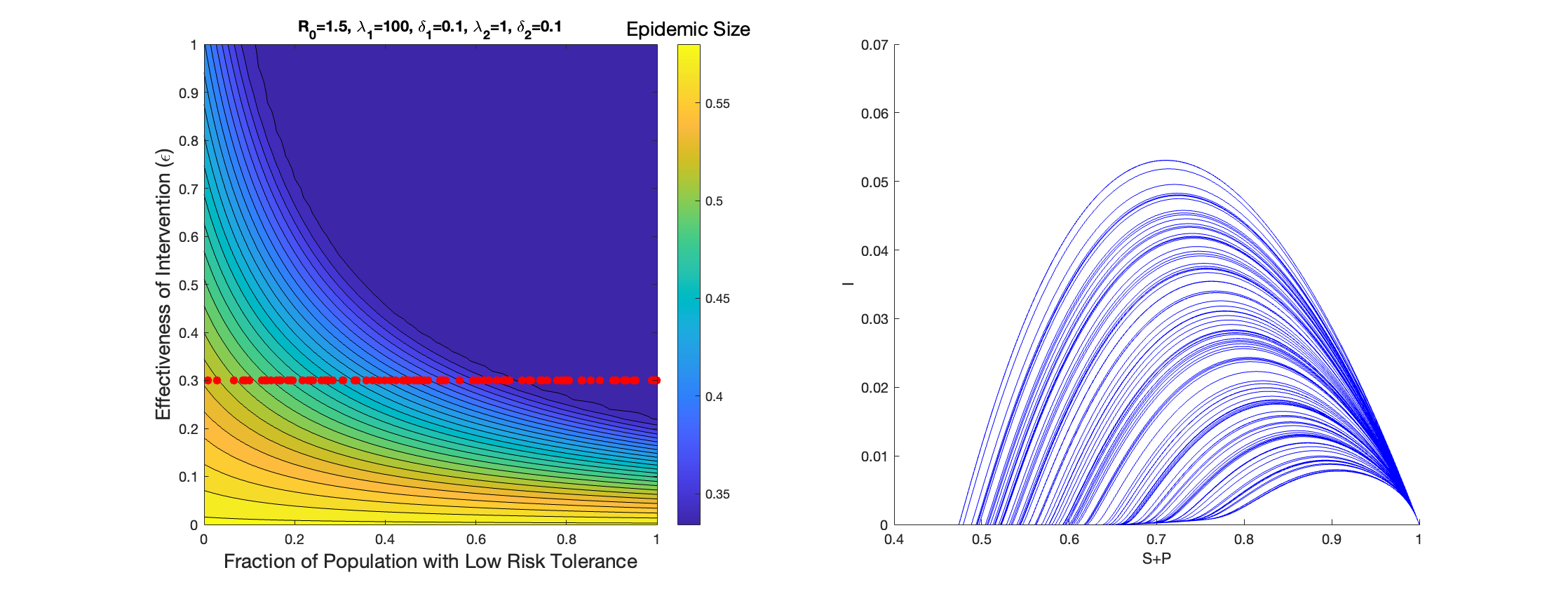}
\caption{Left. Epidemic size as a function of varying the fraction of the population that are low-risk tolerance (i.e. those with higher $\lambda$) when individuals react to incidence rate. Right. Corresponding orbits in the I versus S+P plane for the sampled points in parameter space when the intervention effectiveness has been fixed.} \label{fig:incidence_saturationFixedEpsilon}
\end{figure}

Thus, given the equivalence in results, we have gone with the mathematically simpler and cleaner model based on total number of infected people in the main text. 

\FloatBarrier
\subsubsection*{The Herd Immunity Threshold is Set by a Complex Interplay Between Transmission ($R_0$), Behavior, and Intervention Effectiveness ($\epsilon$).}

While we could make some analytical calculations for the epidemic size in the homogeneous model, the heterogeneous two group case requires a numerical approach to find the plateau region. In general, it is set by a highly nonlinear interaction between the transmission ($R_0$), behavior as determined by the fraction of the population that are risk-averse and risk-taking, and the effectiveness of the intervention ($\epsilon$).

Consider the following progression of figures (Figure \ref{fig:frac}), where in each subsequent figure, the effectiveness of the intervention in blocking transmission ($\epsilon$) is increasing.

\begin{figure*}[t!]
    \centering
    \begin{subfigure}[t]{0.5\textwidth}
        \centering
        \includegraphics[scale=.4]{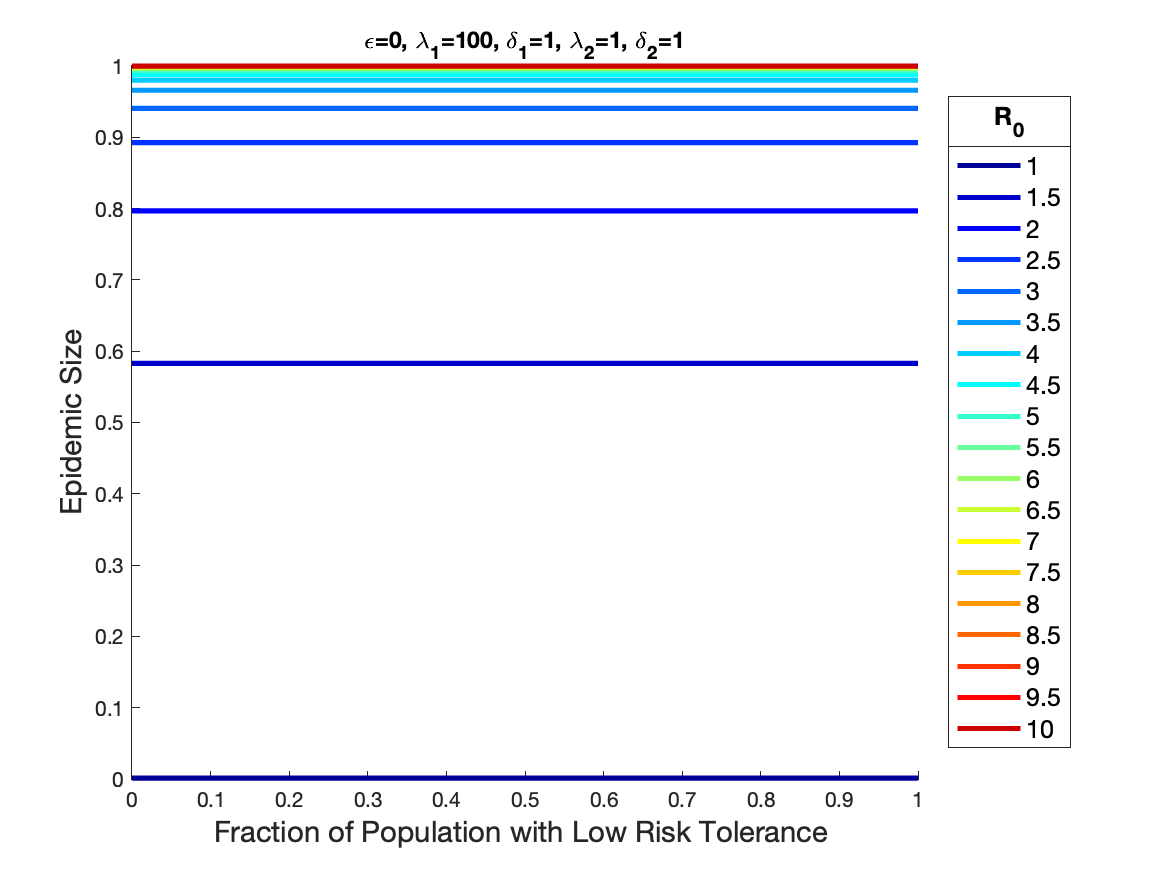}
        \caption{$\epsilon=0$} 
    \end{subfigure}%
    ~ 
    \begin{subfigure}[t]{0.5\textwidth}
        \centering
        \includegraphics[scale=.4]{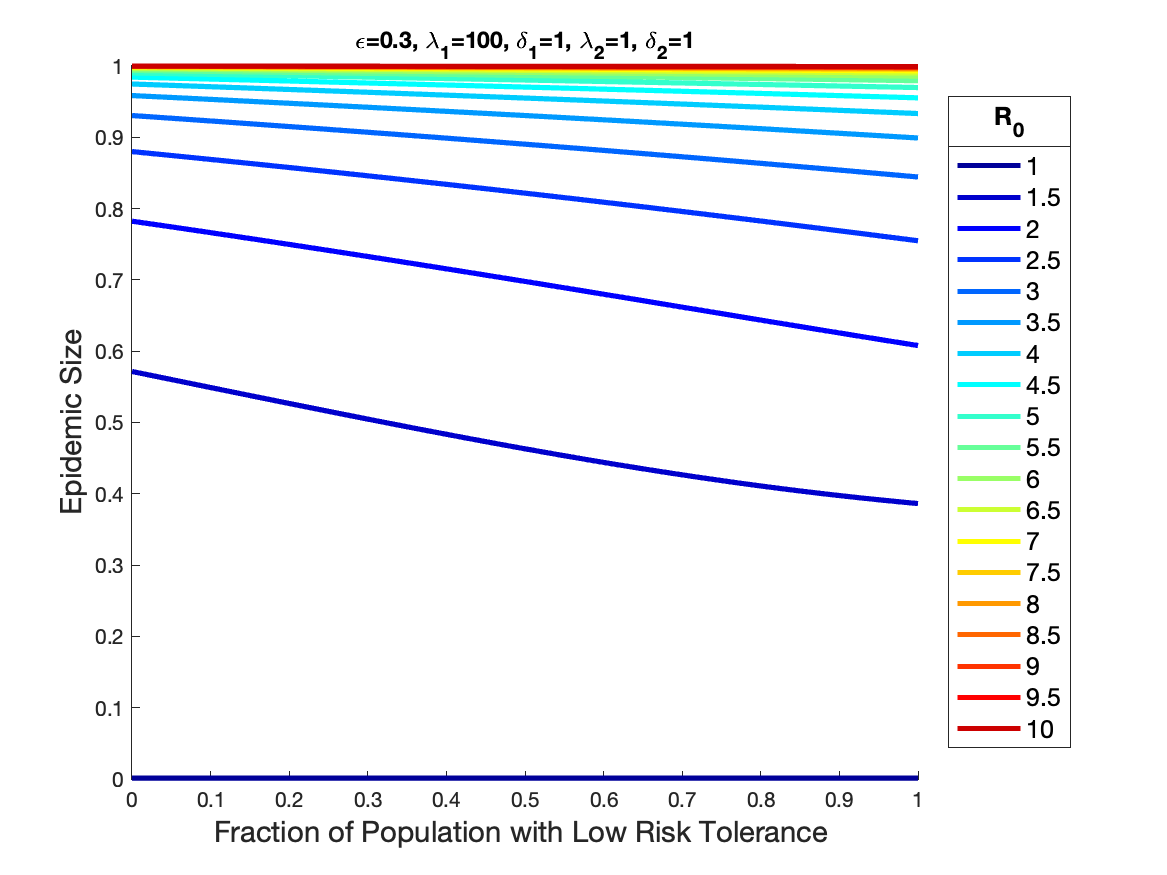}
        \caption{$\epsilon=0.3$}
    \end{subfigure}%
     
    \begin{subfigure}[t]{0.5\textwidth}
        \centering
        \includegraphics[scale=.4]{NPImodel_finalSize_epsilon0.5_lambda1-100_delta1-1_lambda2-1_delta2-1.png}
        \caption{$\epsilon=0.5$}
    \end{subfigure}%
    ~
    \begin{subfigure}[t]{0.5\textwidth}
        \centering
        \includegraphics[scale=.4]{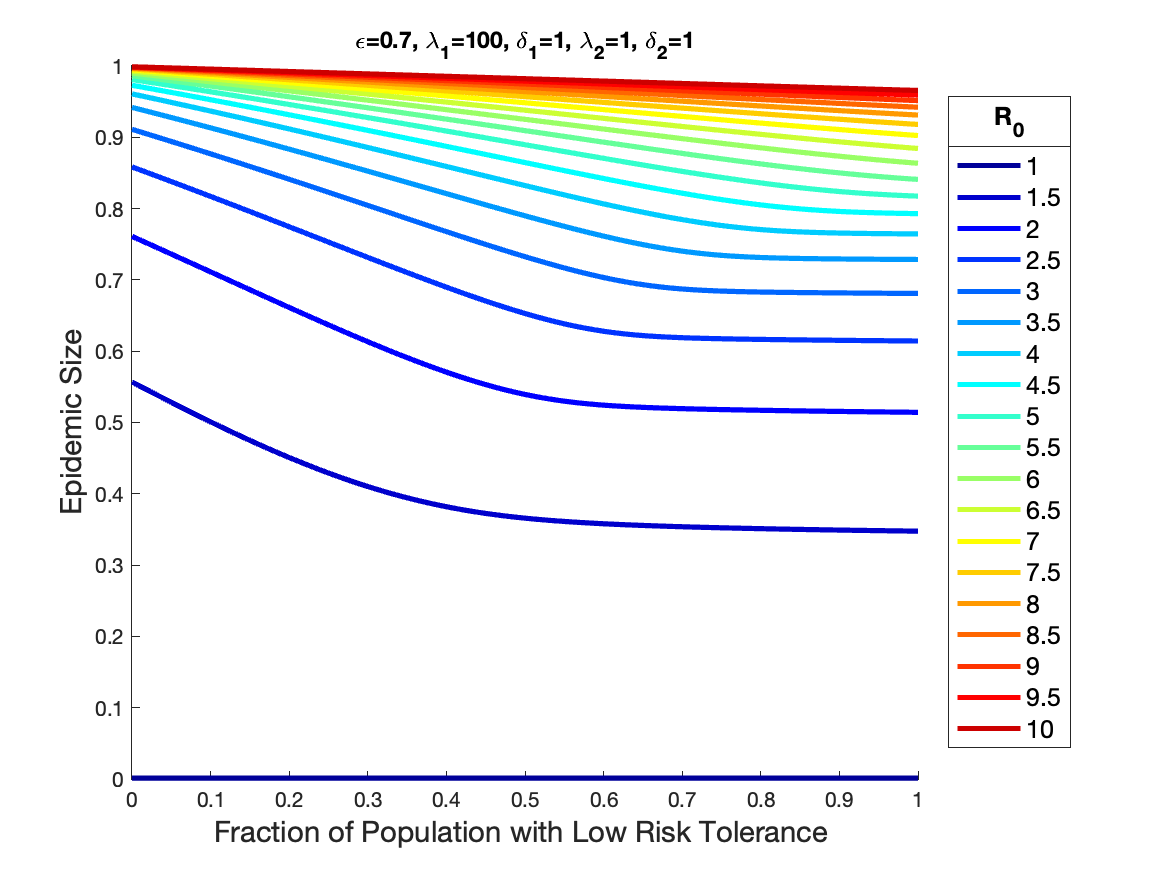}
        \caption{$\epsilon=0.7$}
    \end{subfigure}

    \begin{subfigure}[t]{0.5\textwidth}
        \centering
        \includegraphics[scale=.4]{NPImodel_finalSize_epsilon0.9_lambda1-100_delta1-1_lambda2-1_delta2-1.png}
        \caption{$\epsilon=0.9$}
    \end{subfigure}%
    ~
    \begin{subfigure}[t]{0.5\textwidth}
        \centering
        \includegraphics[scale=.4]{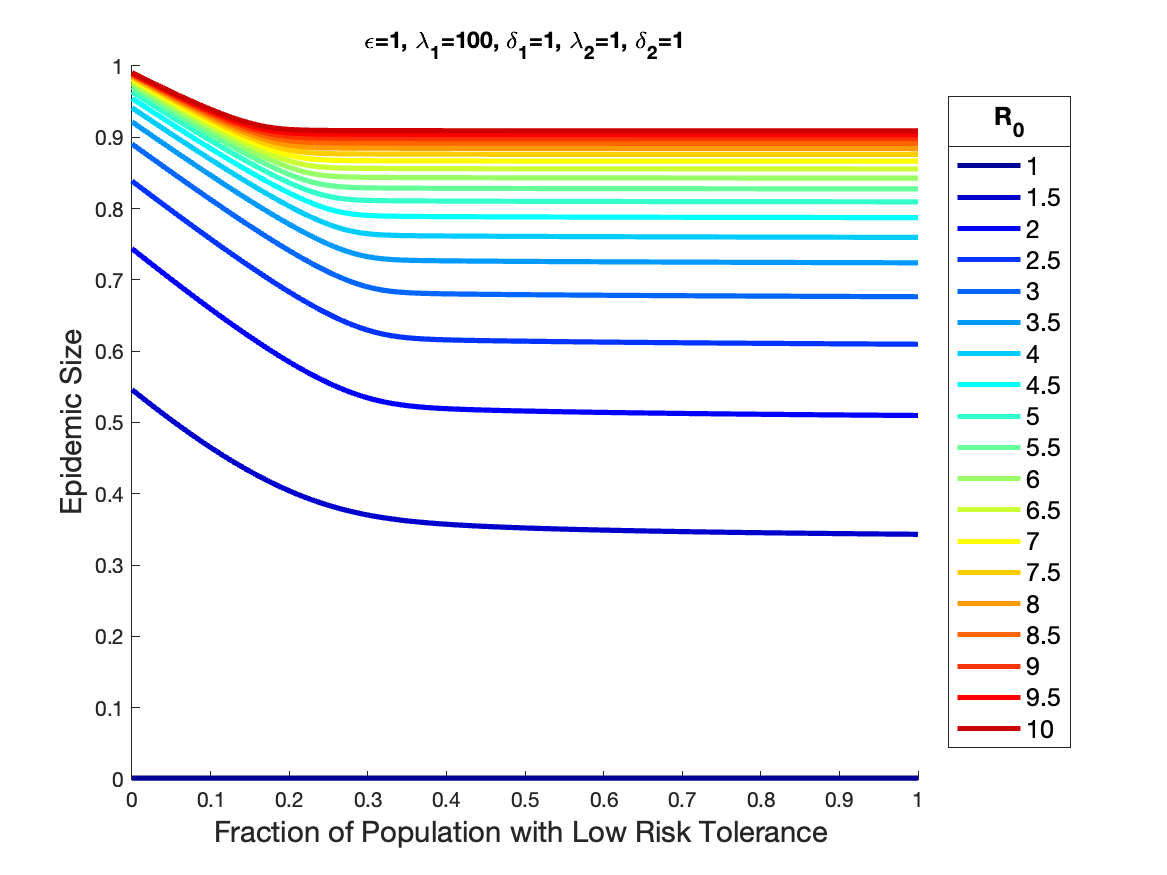}
        \caption{$\epsilon=1$}
    \end{subfigure}
    \caption{Final epidemic size versus fraction of population that are risk-averse ($S_1$) with a progressive increase in intervention effectiveness ($\epsilon$). The other simulation parameters and initial conditions are $\lambda_1 = 100, \delta_1 = 1, \lambda_2 = 1, \delta_2 = 1, I(0) = 10^{-7}, P_1(0)=P_2(0)=0, R(0) = 0$.} \label{fig:frac}
\end{figure*}

\clearpage
\subsubsection*{Comparing Orbits Inside and Outside of the Plateau Region of Herd Immunity}
The following figures sample more orbits for the Figure considered in the main text. 
\begin{figure}[ht]
\centering
\includegraphics[scale=0.2]{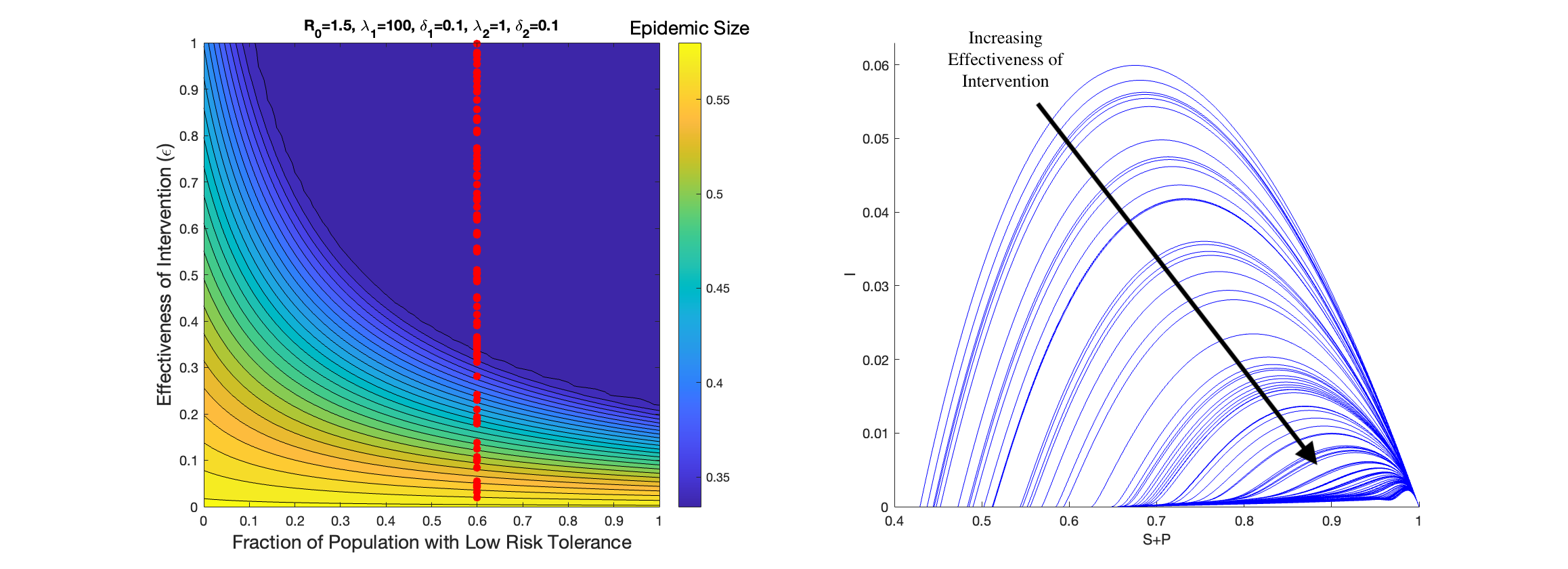}
\caption{Left. Epidemic size as a function of varying the fraction of the population that are low-risk tolerance (i.e. those with higher $\lambda$). Right. Corresponding orbits in the I versus S+P plane for the sampled points in parameter space when the fraction of the population that are low-risk tolerance has been fixed.} \label{fig:saturationFixedX}
\end{figure}

\begin{figure}[ht]
\centering
\includegraphics[scale=0.18]{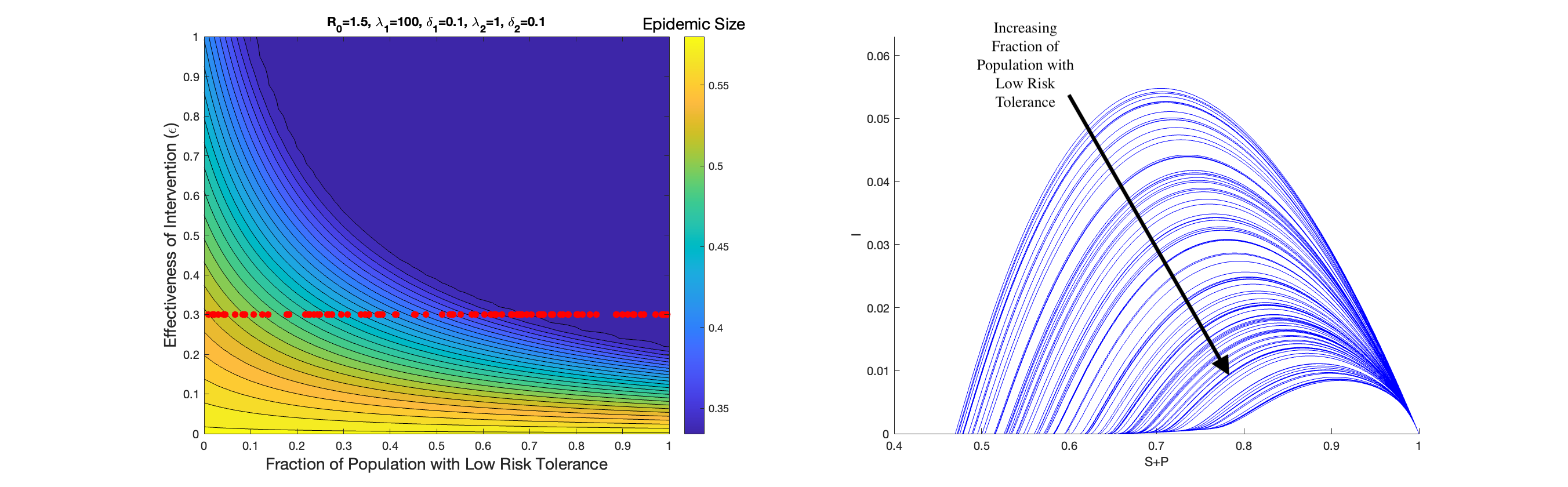}
\caption{Left. Epidemic size as a function of varying the fraction of the population that are low-risk tolerance (i.e. those with higher $\lambda$). Right. Corresponding orbits in the I versus S+P plane for the sampled points in parameter space when the intervention effectiveness has been fixed.} \label{fig:saturationFixedEpsilon}
\end{figure}

\FloatBarrier
\clearpage
\subsubsection*{Proving Underdamped Regime Eliminates Epidemic Overshoot}
In general, the nonlinear feedback between the protected classes and infected individuals make it difficult to make analytical calculations in the full model. Under some simplifications however, we can make some progress. In this section, we derive the epidemic size at which the protection saturates in a restricted case of the homogeneous model ($n=1$). 

Let us consider the homogeneous model in the limit of an intervention with perfect effectiveness (i.e. $\epsilon=1$). We will consider the case where the recovery rate from infection and relaxation rate for interventions are comparable (i.e. $\gamma = \delta$). Since the equation for recovered individuals can be ignored since the population is closed ($S+P+I+R=1$), this reduces the dynamics to the following system:
\begin{align}
\frac{dS}{dt} &= -\beta S I -\lambda S I + \gamma P \label{eqn:simplefirst}\\
\frac{dP}{dt} &=  \lambda S I -\gamma P \label{eqn:simpleP}\\
\frac{dI}{dt} &= \beta S I - \gamma I \label{eqn:simpleI}\\
S(0)&=f,P(0)=0, I(0)=\alpha, R(0)=1-f-\alpha
\end{align}
The initial conditions sets the number of individuals initially susceptible to be $f$, the number of initially infected is assumed to be small $I(0)<<1$, and the remainder of the population is already immune to infection (i.e. recovered). To ensure that the epidemic initially grows in size, we assume that $S(0)>\frac{1}{R_0}$. This follows from comparing the incidence term ($\beta SI$) to the recovery term ($\gamma I$) in \ref{eqn:simpleI}.

To find the asymptotic behavior for $S$, we will attempt to eliminate $P$ from (\ref{eqn:simplefirst}). We start first by seeking an equation that relates the $P$ and $I$ compartments. Consider the following ansatz that considers the difference between the two compartments:
\begin{align}
x = \frac{\beta}{\lambda}P-I \label{eqn:xEqn}
\end{align}
Differentiating this equation with respect to time and using (\ref{eqn:simpleP})-(\ref{eqn:simpleI}) yields:
\begin{align}
\frac{dx}{dt} = \gamma\left(I-\frac{\beta}{\lambda}P\right)
\end{align}
Using (\ref{eqn:xEqn}), this simplifies to $\frac{dx}{dt}=-\gamma x$, which has a critical point at $x=0$. At $x=0$, we obtain that $P=\frac{\lambda}{\beta}I$, indicating a regime where the behavior of $P$ and $I$ scale linearly with each other. Combining this with (\ref{eqn:simplefirst}) yields:
\begin{align}
\frac{dS}{dt} &= \left(-(\beta +\lambda) S + \frac{\gamma\lambda}{\beta}\right)I \label{eqn:modS}
\end{align}

Now that we have an equation that is linear in $I$, we are in a good position to find a final size relationship for the number of susceptibles. To start we take the ratio of (\ref{eqn:simpleI}) and (\ref{eqn:modS}).
\begin{align}
\frac{dI}{dS} &= \frac{\beta SI-\gamma I}{\left(-(\beta +\lambda) S + \frac{\gamma\lambda}{\beta}\right)I} \label{eqn:dIdS}
\end{align}
Using the partial fractions $\frac{-\beta}{\beta+\lambda}$ and $\frac{\frac{\gamma\beta}{\beta+\lambda}}{((\beta +\lambda) S - \frac{\gamma\lambda}{\beta})}$, we get upon indefinite integration of (\ref{eqn:dIdS}) that \newline $k = \frac{\beta}{\beta+\lambda}\left(\frac{\gamma}{\beta+\lambda}\text{ln}\left((\beta +\lambda) S - \frac{\gamma\lambda}{\beta}\right)-S\right)-I$, where $k$ is a constant that holds throughout the trajectory of the dynamics. Thus, considering the values of $S$ and $I$ at the beginning of the epidemic ($t=0$) and the end of the epidemic ($t=\infty$) and using the conditions that $I(\infty)=0$ and $I(0) \approx 0$ yields the following transcendental equation for the final epidemic size. 
\begin{align}
\frac{\beta}{\beta+\lambda}\left(\frac{\gamma}{\beta+\lambda}\text{ln}\left((\beta +\lambda) S(0) - \frac{\gamma\lambda}{\beta}\right)-S(0)\right)&= \frac{\beta}{\beta+\lambda}\left(\frac{\gamma}{\beta+\lambda}\text{ln}\left((\beta +\lambda) S(\infty) - \frac{\gamma\lambda}{\beta}\right)-S(\infty)\right)\label{eqn:finalSize}
\end{align}

Since $S(0) > \frac{1}{R_0}$, then the argument of the logarithm on the left hand side must be positive, and subsequently the left hand side evaluates to a real number. Due to the equality, the right hand side must also evaluate to a real number, implying the argument of the logarithm on the right hand side must also be positive. Positivity implies the following inequality for the lower bound for the final number of susceptibles:
\begin{align}
S(\infty) > \frac{\gamma\lambda}{\beta(\beta +\lambda)} \label{eqn:sInequality}
\end{align}

An upper bound can be given by simply noting that in the long time limit, the recovery term ($\gamma I$) must be at least as large as the incidence term ($\beta S(\infty)I$) in (\ref{eqn:simpleI}), otherwise the epidemic would still be growing. This implies $S(\infty)\leq \frac{1}{R_0}$.

To summarize, when the interventions are perfectly effective, the rate of relaxation from the protected class is equal to the rate of recovery from infection, and the number in the protected class scales linearly with the number of infected, then the final fraction of susceptibles is bounded as follows:
\begin{align}
\frac{\lambda}{R_0(\beta +\lambda)}<S(\infty)\leq \frac{1}{R_0}
\end{align} 

We see in the parameter limit of when the adoption rate of interventions is very fast compared to the transmission rate (i.e. $\lambda >> \beta$), that the lower bound reduces to $\frac{1}{R_0}$. Since both bounds now coincide, then $S_{\infty}$ must equal that value. Interestingly this corresponds to the herd immunity threshold of the standard SIR model. As the overshoot is the excess number of cases beyond the herd immunity threshold, we see that in this parameter limit there is no overshoot.

This analysis for the homogeneous case also carries over to the heterogeneous case for two groups when the adoption rate between the two groups is significantly different (i.e. $\lambda_1 >> \lambda_2$). This results in a separation of time scales in which the faster adopters quickly transition to the protected state and can essentially be treated as immune over the course of the remaining epidemic over the slow adopters. This amounts to effectively reducing the dynamics to the homogeneous model considered here where $f$ and $1-f$ fractions of the population in the susceptible and recovered respectively correspond to the fraction of the population in the slow adopter ($\lambda_2$) and fast adopter groups ($\lambda_1$).

\clearpage
\subsubsection*{Heterogeneity in Risk Tolerance through Arithmetic Averaging}
There are two ways for calculating the difference (heterogeneity) in adoption rates, either through geometric or arithmetic averaging. Both can be justified, and we presented the geometric formulation in the main text. We find that the arithmetic formulation gives qualitative similar results under a suitable parameter shift (Figure \ref{fig:meanVariance_arithmetic}). 
\begin{figure}[ht]
\centering
\includegraphics[scale=0.15]{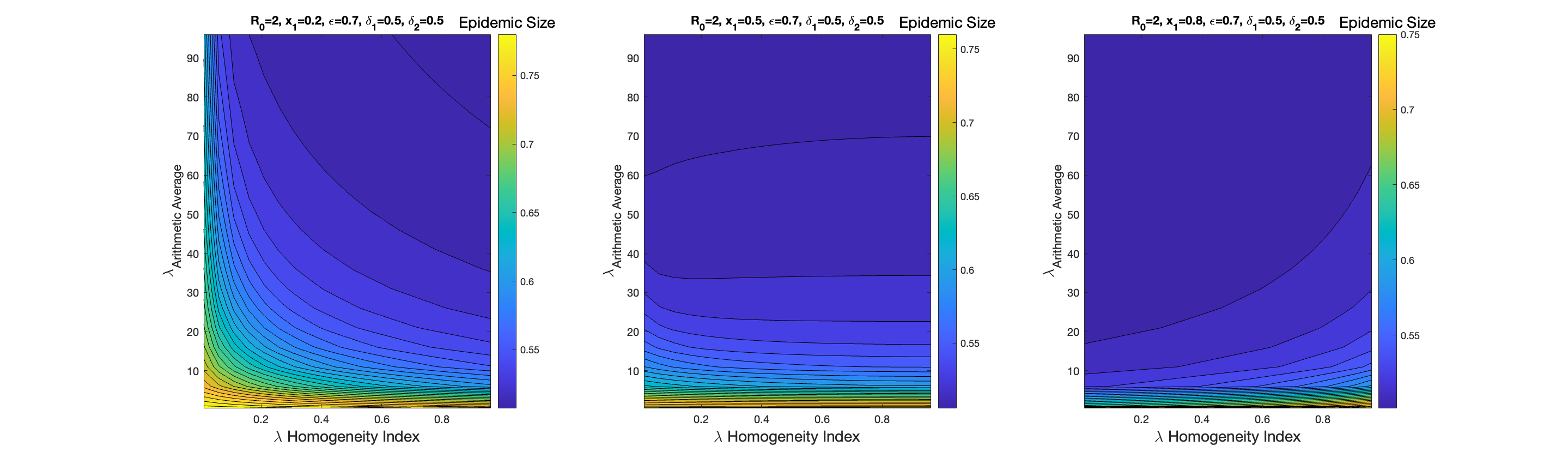}
\caption{Epidemic size under differing levels of heterogeneity in the adoption rate for interventions. The mean adoption rate of the two groups (i.e. arithmetic average of $\lambda_1, \lambda_2$) is compared to the difference in the two adoption rates as parameterized by a homogeneity index (see Methods for definition). $Left$ is when the fraction of the population with low risk tolerance ($x_1$) is 0.2, $center$ is when $x_1=0.5$, $right$ is when $x_1=0.8$.} \label{fig:meanVariance_arithmetic}
\end{figure}

\FloatBarrier

\subsection*{Code to Generate Figures}
Code executed in MATLAB R2023a.
\UseRawInputEncoding

\begin{lstlisting}[
frame=single,
numbers=left,
style=Matlab-Pyglike]
%% %%%%%%%%%%%%%%%%%%%%%%%%%%%%%%%%%%%%%%%%%%%%% Functions %%%%%%%%%%%%%%%%%%%%%%%%%%%%%%%%%%%%%%%%%%%%%%% 

function totalRecovered = plotSIRDynamics(susceptibleFrac, R0, transmissionReduction, PlotOption, gamma, lambda1, delta1, lambda2, delta2, modelType)

beta = R0*gamma;   

% Initial conditions
initialInfected = 0.00000001;
initialSusceptible1 = susceptibleFrac;%(1-initialInfected)/2;
initialSusceptible2 = 1-initialInfected-initialSusceptible1;
initialMasked1 = 0;
initialMasked2 = 0;
initialRecovered = 0;

% Time vector
tspan = [0 10000];

% SIR ODE system

%% Reversible w/ total level response
sirODE_level = @(t, y) [
    -beta*y(1)*y(5) - lambda1*y(1)*y(5) + delta1*y(2);
    -beta*(1-transmissionReduction)*y(2)*y(5) - delta1*y(2) + lambda1*y(1)*y(5);
    -beta*y(3)*y(5) - lambda2*y(3)*y(5) + delta2*y(4);
    -beta*(1-transmissionReduction)*y(4)*y(5) - delta2*y(4) + lambda2*y(3)*y(5);
    beta*(y(1)+y(3))*y(5) + beta*(1-transmissionReduction)*(y(2)+y(4))*y(5) - gamma*y(5);
    gamma * y(5)];

%% Reversible w/ incidence rate response

sirODE_incidence = @(t, y) [
    -beta*y(1)*y(5) - lambda1*y(1)*(beta*(y(1)+y(3))*y(5) + beta*(1-transmissionReduction)*(y(2)+y(4))*y(5)) + delta1*y(2);
    -beta*(1-transmissionReduction)*y(2)*y(5) - delta1*y(2) + lambda1*y(1)*(beta*(y(1)+y(3))*y(5) + beta*(1-transmissionReduction)*(y(2)+y(4))*y(5));
    -beta*y(3)*y(5) - lambda2*y(3)*(beta*(y(1)+y(3))*y(5) + beta*(1-transmissionReduction)*(y(2)+y(4))*y(5)) + delta2*y(4);
    -beta*(1-transmissionReduction)*y(4)*y(5) - delta2*y(4) + lambda2*y(3)*(beta*(y(1)+y(3))*y(5) + beta*(1-transmissionReduction)*(y(2)+y(4))*y(5));
    beta*(y(1)+y(3))*y(5) + beta*(1-transmissionReduction)*(y(2)+y(4))*y(5) - gamma*y(5);
    gamma * y(5)];

% Solve ODE
if modelType == 1
    [t, y] = ode89(sirODE_level, tspan, [initialSusceptible1; initialMasked1; initialSusceptible2; initialMasked2; initialInfected; initialRecovered]);
else
    [t, y] = ode89(sirODE_incidence, tspan, [initialSusceptible1; initialMasked1; initialSusceptible2; initialMasked2; initialInfected; initialRecovered]);
end

if PlotOption == 1
%    Plot results
    figure;
    plot(t, y(:, 1), 'k-', 'LineWidth', 2, 'DisplayName', 'Susceptible - Type 1');
    hold on;
    plot(t, y(:, 3), 'b-', 'LineWidth', 2, 'DisplayName', 'Susceptible - Type 2');
    plot(t, y(:, 2), 'g-', 'LineWidth', 2, 'DisplayName', 'Protected - Type 1');
    plot(t, y(:, 4), 'm-', 'LineWidth', 2, 'DisplayName', 'Protected - Type 2');
    plot(t, y(:, 5), 'r-', 'LineWidth', 2, 'DisplayName', 'Infected');
    plot(t, y(:, 6), 'b-', 'LineWidth', 2, 'DisplayName', 'Recovered');
    plot(t, y(:, 1)+y(:, 2)+y(:, 3)+y(:, 4), 'k.', 'LineWidth', 2, 'DisplayName', 'Total Susceptible + Protected');
    xlabel('Time');
    ylabel('Proportion of Population');
    
    ylim([0 1])
    title(['\beta=',num2str(beta),', \gamma=',num2str(gamma),', \epsilon=',num2str(transmissionReduction), ', \lambda_1=',num2str(lambda1),', \delta_1=',num2str(delta1),', \lambda_2=',num2str(lambda2),', \delta_2=',num2str(delta2),', x_{S_1}=',num2str(susceptibleFrac)]);
    legend('Location', 'eastoutside');
    grid on;
    hold off;

end

if PlotOption == 2
    hold on
    plot(y(:, 1)+y(:,2)+y(:, 3)+y(:,4), y(:,5), 'b-')
    xlabel('S+P')
    ylabel('I')
    ylim([0 .15])
end

totalRecovered = y(end,6);
end

%% %%%%%%%%%%%%%%%%%%%%%%%%%%%%%%%%%%%%%%%%%%%%% Script %%%%%%%%%%%%%%%%%%%%%%%%%%%%%%%%%%%%%%%%%%%%%%% 

%%%%%%%%%%%%%%%%%%%%%%%%%%%% Figure 2 %%%%%%%%%%%%%%%%%%%%%%%%%%%% 

%%%%% SIR model parameters
R0 = 3; % Basic reproduction number
gamma = 1;   % Recovery rate
beta = R0*gamma;   % Transmission rate of unmasked susceptibles
transmissionReduction = 0.8; % Effectiveness of intervention (\epsilon)
lambda1 = 10; % Rate parameter for unprotected susceptibles to adopt intervention
delta1 = .01; % Rate parameter for protected individuals to remove intervention (mask)

PlotOption = 1; % Plotting parameter. If 1, then plot graph, otherwise none.

% Initial conditions
initialInfected = 0.000001; % Seed fraction of population that are initially infected
initialSusceptible1 = 1-initialInfected;
initialMasked1 = 0;
initialRecovered = 0;

% Time vector
tspan = [0 1500];

% SIR ODE system
%% Reversible w/ total level response
sirODE_level = @(t, y) [
    -beta*y(1)*y(3) - lambda1*y(1)*y(3) + delta1*y(2);
    -beta*(1-transmissionReduction)*y(2)*y(3) - delta1*y(2) + lambda1*y(1)*y(3);
    beta*y(1)*y(3) + beta*(1-transmissionReduction)*y(2)*y(3) - gamma*y(3);
    gamma * y(3)];

% Solve ODE
[t, y] = ode89(sirODE_level, tspan, [initialSusceptible1; initialMasked1; initialInfected; initialRecovered]);

if PlotOption == 1
    % Plot results
    figure;
    plot(t, y(:, 1), 'k-', 'LineWidth', 2, 'DisplayName', 'Susceptible - Type 1');
    hold on;
    plot(t, y(:, 2), 'g-', 'LineWidth', 2, 'DisplayName', 'Protected - Type 1');
    plot(t, y(:, 3), 'r-', 'LineWidth', 2, 'DisplayName', 'Infected');
    plot(t, y(:, 4), 'b-', 'LineWidth', 2, 'DisplayName', 'Recovered');
    plot(t, y(:, 1)+y(:, 2), 'c-', 'LineWidth', 2, 'DisplayName', 'Total Susceptible + Protected');
    xlabel('Time');
    ylabel('Proportion of Population');
    ylim([0 1])
    title(['\beta=',num2str(beta),', \gamma=',num2str(gamma),', \epsilon=',num2str(transmissionReduction), ', \lambda_1=',num2str(lambda1),', \delta_1=',num2str(delta1)]);
    legend('Location', 'eastoutside');
    grid on;
    hold off;
% 
%     saveas(gcf,['NPImodel_','beta',num2str(beta),'_gamma',num2str(gamma),'_epsilon',num2str(transmissionReduction), '_lambda1-',num2str(lambda1),'_delta1-',num2str(delta1),'.png'])

    figure
    hold on
    plot3(y(:, 1),y(:,2), y(:,3), 'b-')
    xlabel('S')
    ylabel('P')
    zlabel('I')
end

totalRecovered = y(end,4);


%%%%%%%%%%%%%%%%%%%%%%%%%%%% Figure 3 %%%%%%%%%%%%%%%%%%%%%%%%%%%% 
susceptFracVec = 0:0.02:1;
R0vec = 2;
colorVec = jet(length(R0vec));
transmissionReduction = 0:0.02:1;

% SIR model parameters
gamma = 1;   % Recovery rate
lambda1 = 100; % Rate of susceptibles with lower tolerance to switch to masked class
delta1 = .1; % Rate of masked class with lower tolerance for adoption to remove intervention (mask) to return to susceptible class
lambda2 = 1; % Rate of susceptibles with higher tolerance to switch to masked class
delta2 = .1; % Rate of masked class with higher tolerance for adoption to remove intervention (mask) to return to susceptible class

recoveredFracVec = zeros(length(susceptFracVec),length(transmissionReduction),length(R0vec));
xVec = zeros(length(susceptFracVec),length(transmissionReduction),length(R0vec));
effectivenessVec = zeros(length(susceptFracVec),length(transmissionReduction),length(R0vec));

for z = 1:length(R0vec)
    for y = 1:length(transmissionReduction)
        for x = 1:length(susceptFracVec)
            recoveredFracVec(x,y,z) = plotSIRDynamics(susceptFracVec(x), R0vec(z),transmissionReduction(y), 0, gamma, lambda1, delta1, lambda2, delta2, 1);
            xVec(x,y,z) = susceptFracVec(x);
            effectivenessVec(x,y,z) = transmissionReduction(y);
        end
    end
end

figure

subplot(1,2,1)
[x_new, e_new] = meshgrid(susceptFracVec, transmissionReduction);
contourf(x_new,e_new,recoveredFracVec(:,:,z)', 'LevelStep', 0.01)
xlabel('Fraction of Population with Low Risk Tolerance', 'FontSize', 14)
ylabel('Effectiveness of Intervention (\epsilon)', 'FontSize', 14)
h = colorbar;
title(h, 'Epidemic Size', 'FontSize', 14)
title(['R_0=', num2str(R0vec(z)), ', \lambda_1=', num2str(lambda1), ', \delta_1=', num2str(delta1), ', \lambda_2=', num2str(lambda2), ', \delta_2=', num2str(delta2)]);

hold on
points_x = [0.05, 0.2,.9, 0.5, .95,  .75];
points_y = [0.8, 0.1, .2,0.9, .95,  .5];

scatter(points_x, points_y, 'r', 'filled')

labels = {'A', 'B', 'C', 'D', 'E', 'F'};
text(points_x, points_y, labels, 'Color', 'red', 'VerticalAlignment', 'bottom', 'HorizontalAlignment', 'right')

subplot(1,2,2)

hold on
for a = 1:length(points_x)
    plotSIRDynamics(points_x(a), R0vec(z),points_y(a), 2, gamma, lambda1, delta1, lambda2, delta2, 1);
end

points_x = [0.45,0.4, .36, .85,  .95, 0.72];
points_y = [0.025,0.13, 0.02,.01,  .002, 0.02];

scatter(points_x, points_y, 'Marker', 'none')
labels = {'A', 'B', 'C', 'D', 'E', 'F'};
text(points_x, points_y, labels, 'Color', 'red', 'VerticalAlignment', 'bottom', 'HorizontalAlignment', 'right')


%%%%%%%%%%%%%%%%%%%%%%%%%%%% Figure 4 and S3 %%%%%%%%%%%%%%%%%%%%%%%%%%%%
% Repeat for different parameter conditions to generate each panel

susceptFracVec = 0:0.01:1; 
R0vec = 1:0.5:10;
colorVec = jet(length(R0vec));
transmissionReduction = 1;

% SIR model parameters
gamma = 1;   % Recovery rate
lambda1 = 100; % Rate of susceptibles with lower tolerance to switch to masked class
delta1 = .1; % Rate of masked class with lower tolerance for adoption to remove intervention (mask) to return to susceptible class
lambda2 = 1; % Rate of susceptibles with higher tolerance to switch to masked class
delta2 = .1; % Rate of masked class with higher tolerance for adoption to remove intervention (mask) to return to susceptible class


recoveredFracVec = zeros(length(R0vec),length(susceptFracVec));
figure
hold on
for y = 1:length(R0vec)
    for x = 1:length(susceptFracVec)
        recoveredFracVec(y, x) = plotSIRDynamics(susceptFracVec(x), R0vec(y),transmissionReduction, 0, gamma, lambda1, delta1, lambda2, delta2, 1);
    end
    hold on
    plot(susceptFracVec, recoveredFracVec(y,:), 'Color', colorVec(y,:), 'LineWidth',2.5)
end

xlabel('Fraction of Population that are Risk Averse', 'FontSize', 14)
ylabel('Epidemic Size', 'FontSize', 14)
title(['\epsilon=',num2str(transmissionReduction), ', \lambda_1=',num2str(lambda1),', \delta_1=',num2str(delta1),', \lambda_2=',num2str(lambda2),', \delta_2=',num2str(delta2)]);
hLegend = legend(string(R0vec), 'Location', 'eastoutside', 'FontSize', 12); 
title(hLegend, 'R_0')

%%%%%%%%%%%%%%%%%%%%%%%%%%%% Figure 5 %%%%%%%%%%%%%%%%%%%%%%%%%%%%
 % Repeat for different parameter conditions to generate each panel

figure
susceptFracVec = 0.2;
R0vec = 2;
colorVec = jet(length(R0vec));
transmissionReduction = .7;

% SIR model parameters
cVec = [0.01:0.01:1];
lambdaAvg = [0.5, 1:1:100];
gamma = 1;   % Recovery rate
delta1 = .5; % Rate of masked class with lower tolerance for adoption to remove intervention (mask) to return to susceptible class
delta2 = .5; % Rate of masked class with higher tolerance for adoption to remove intervention (mask) to return to susceptible class

recoveredFracVec = zeros(length(cVec),length(lambdaAvg));

for x = 1:length(cVec)
    for y = 1:length(lambdaAvg)
        recoveredFracVec(x,y) = plotSIRDynamics(susceptFracVec, R0vec, transmissionReduction, 0, gamma, lambdaAvg(y)/(cVec(x)^(susceptFracVec/(1-susceptFracVec))), delta1, cVec(x)*lambdaAvg(y), delta2, 1);
    end
end

[c_new, l_new] = meshgrid(cVec, lambdaAvg);
contourf(c_new,l_new,recoveredFracVec', 'LevelStep', 0.01)
xlabel('\lambda Homogeneity Index', 'FontSize', 14)
ylabel('\lambda_{Geometric Average}', 'FontSize', 14)
h = colorbar;
title(h, 'Epidemic Size', 'FontSize', 14)
title(['R_0=', num2str(R0vec), ', x_1=', num2str(susceptFracVec), ', \epsilon=', num2str(transmissionReduction), ', \delta_1=', num2str(delta1),', \delta_2=', num2str(delta2)]);

%%%%%%%%%%%%%%%%%%%%%%%%%%%% Figure 6 %%%%%%%%%%%%%%%%%%%%%%%%%%%%

figure
subplot(1,2,1)

susceptFracVec = 0:0.01:1;
R0vec = 1:0.5:7;
colorVec = jet(length(R0vec));
transmissionReduction = 1;

% SIR model parameters
gamma = 1;   % Recovery rate
lambda1 = 10; % Rate of susceptibles with lower tolerance to switch to masked class
delta1 = .1; % Rate of masked class with lower tolerance for adoption to remove intervention (mask) to return to susceptible class
lambda2 = .5; % Rate of susceptibles with higher tolerance to switch to masked class
delta2 = .1; % Rate of masked class with higher tolerance for adoption to remove intervention (mask) to return to susceptible class


recoveredFracVec = zeros(length(R0vec),length(susceptFracVec));
figure
hold on
for y = 1:length(R0vec)
    for x = 1:length(susceptFracVec)
        recoveredFracVec(y, x) = plotSIRDynamics(susceptFracVec(x), R0vec(y),transmissionReduction, 0, gamma, lambda1, delta1, lambda2, delta2, 1);
    end
    hold on
    plot(susceptFracVec, recoveredFracVec(y,:), 'Color', colorVec(y,:), 'LineWidth',2.5)
end

xlabel('Fraction of Population that are Risk Averse', 'FontSize', 14)
ylabel('Epidemic Size', 'FontSize', 14)
title(['\epsilon=',num2str(transmissionReduction), ', \lambda_1=',num2str(lambda1),', \delta_1=',num2str(delta1),', \lambda_2=',num2str(lambda2),', \delta_2=',num2str(delta2)]);
hLegend = legend(string(R0vec), 'Location', 'eastoutside', 'FontSize', 12); 
title(hLegend, 'R_0')

susceptFracVec = 0:0.02:1;
R0vec = 2.5; % Only pick one R0 to show the surface plot for. Use other plot to pick the value of interest
colorVec = jet(length(R0vec));
transmissionReduction = 0:0.02:1;

recoveredFracVec = zeros(length(susceptFracVec),length(transmissionReduction),length(R0vec));
xVec = zeros(length(susceptFracVec),length(transmissionReduction),length(R0vec));
effectivenessVec = zeros(length(susceptFracVec),length(transmissionReduction),length(R0vec));

for z = 1:length(R0vec)
    for y = 1:length(transmissionReduction)
        for x = 1:length(susceptFracVec)
            recoveredFracVec(x,y,z) = plotSIRDynamics(susceptFracVec(x), R0vec(z),transmissionReduction(y), 0, gamma, lambda1, delta1, lambda2, delta2, 1);
            xVec(x,y,z) = susceptFracVec(x);
            effectivenessVec(x,y,z) = transmissionReduction(y);
        end
    end
end

[X,Y,Z] = meshgrid(susceptFracVec,transmissionReduction,R0vec);

subplot(1,2,2)
for z = 1:length(R0vec)
    surf(X(:,:,z), Y(:,:,z), recoveredFracVec(:,:,z), recoveredFracVec(:,:,z), 'FaceAlpha', 0.8, 'EdgeColor', 'interp');
    hold on
end

xlabel('Fraction of Population that are Risk-Averse', 'FontSize', 14)
ylabel('Effectiveness of Intervention (\epsilon)', 'FontSize', 14)
zlabel('Epidemic Size', 'FontSize', 14)
title(['\lambda_1=', num2str(lambda1), ', \delta_1=', num2str(delta1), ', \lambda_2=', num2str(lambda2), ', \delta_2=', num2str(delta2)]);


%%%%%%%%%%%%%%%%%%%%%%%%%%%% Figure 7 %%%%%%%%%%%%%%%%%%%%%%%%%%%%

maskPolicyDataVsCases = readtable('maskPolicyDataVsCases.xlsx');

dates = datetime(maskPolicyDataVsCases.Date, 'InputFormat', 'your_date_format'); % Replace 'your_date_format' with the actual format
policy = maskPolicyDataVsCases.MaskPolicyLevel;
cases = maskPolicyDataVsCases.ConfirmedNewCases;

figure; 

% Plot the first variable on the left y-axis
yyaxis left;
plot(dates, policy);
ylabel('Mask Policy Level');
grid off; 
ylim([0, 5]); % Uncomment and adjust if you need specific limits for the left y-axis

% Plot the second variable on the right y-axis
yyaxis right;
plot(dates, cases);
ylabel('Confirmed New Cases in USA');


datetick('x', 'yyyy-mm-dd');
xlabel('Date');
legend('Mask Policy Level', 'Confirmed New Cases');


%%%%%%%%%%%%%%%%%%%%%%%%%%%% Figure S1 %%%%%%%%%%%%%%%%%%%%%%%%%%%%

susceptFracVec = 0:0.02:1;
R0vec = 2;
colorVec = jet(length(R0vec));
transmissionReduction = 0:0.02:1;

% SIR model parameters
gamma = 1;   % Recovery rate
lambda1 = 100; % Rate of susceptibles with lower tolerance to switch to masked class
delta1 = .1; % Rate of masked class with lower tolerance for adoption to remove intervention (mask) to return to susceptible class
lambda2 = 1; % Rate of susceptibles with higher tolerance to switch to masked class
delta2 = .1; % Rate of masked class with higher tolerance for adoption to remove intervention (mask) to return to susceptible class

recoveredFracVec = zeros(length(susceptFracVec),length(transmissionReduction),length(R0vec));
xVec = zeros(length(susceptFracVec),length(transmissionReduction),length(R0vec));
effectivenessVec = zeros(length(susceptFracVec),length(transmissionReduction),length(R0vec));

for z = 1:length(R0vec)
    for y = 1:length(transmissionReduction)
        for x = 1:length(susceptFracVec)
            recoveredFracVec(x,y,z) = plotSIRDynamics(susceptFracVec(x), R0vec(z),transmissionReduction(y), 0, gamma, lambda1, delta1, lambda2, delta2, 0);  % individuals adopt based on incidence
            xVec(x,y,z) = susceptFracVec(x);
            effectivenessVec(x,y,z) = transmissionReduction(y);
        end
    end
end


figure

subplot(1,2,1)
[x_new, e_new] = meshgrid(susceptFracVec, transmissionReduction);
contourf(x_new,e_new,recoveredFracVec(:,:,z)', 'LevelStep', 0.01)
xlabel('Fraction of Population with Low Risk Tolerance', 'FontSize', 14)
ylabel('Effectiveness of Intervention (\epsilon)', 'FontSize', 14)
h = colorbar;
title(h, 'Epidemic Size', 'FontSize', 14)
title(['R_0=', num2str(R0vec(z)), ', \lambda_1=', num2str(lambda1), ', \delta_1=', num2str(delta1), ', \lambda_2=', num2str(lambda2), ', \delta_2=', num2str(delta2)]);

hold on
points_x = 0.6*ones(100,1);
points_y = rand(100,1);

scatter(points_x, points_y, 'r', 'filled')

subplot(1,2,2)

for a = 1:length(points_x)
    plotSIRDynamics(points_x(a), R0vec(z),points_y(a), 1, gamma, lambda1, delta1, lambda2, delta2, 0);  % individuals adopt based on incidence
end


%%%%%%%%%%%%%%%%%%%%%%%%%%%% Figure S2 %%%%%%%%%%%%%%%%%%%%%%%%%%%%

susceptFracVec = 0:0.02:1;
R0vec = 2;
colorVec = jet(length(R0vec));
transmissionReduction = 0:0.02:1;

% SIR model parameters
gamma = 1;   % Recovery rate
lambda1 = 100; % Rate of susceptibles with lower tolerance to switch to masked class
delta1 = .1; % Rate of masked class with lower tolerance for adoption to remove intervention (mask) to return to susceptible class
lambda2 = 1; % Rate of susceptibles with higher tolerance to switch to masked class
delta2 = .1; % Rate of masked class with higher tolerance for adoption to remove intervention (mask) to return to susceptible class

recoveredFracVec = zeros(length(susceptFracVec),length(transmissionReduction),length(R0vec));
xVec = zeros(length(susceptFracVec),length(transmissionReduction),length(R0vec));
effectivenessVec = zeros(length(susceptFracVec),length(transmissionReduction),length(R0vec));

for z = 1:length(R0vec)
    for y = 1:length(transmissionReduction)
        for x = 1:length(susceptFracVec)
            recoveredFracVec(x,y,z) = plotSIRDynamics(susceptFracVec(x), R0vec(z),transmissionReduction(y), 0, gamma, lambda1, delta1, lambda2, delta2, 0);  % individuals adopt based on incidence
            xVec(x,y,z) = susceptFracVec(x);
            effectivenessVec(x,y,z) = transmissionReduction(y);
        end
    end
end


figure

subplot(1,2,1)
[x_new, e_new] = meshgrid(susceptFracVec, transmissionReduction);
contourf(x_new,e_new,recoveredFracVec(:,:,z)', 'LevelStep', 0.01)
xlabel('Fraction of Population with Low Risk Tolerance', 'FontSize', 14)
ylabel('Effectiveness of Intervention (\epsilon)', 'FontSize', 14)
h = colorbar;
title(h, 'Epidemic Size', 'FontSize', 14)
title(['R_0=', num2str(R0vec(z)), ', \lambda_1=', num2str(lambda1), ', \delta_1=', num2str(delta1), ', \lambda_2=', num2str(lambda2), ', \delta_2=', num2str(delta2)]);

hold on
points_x = rand(100,1);
points_y = 0.3*ones(100,1);

scatter(points_x, points_y, 'r', 'filled')

subplot(1,2,2)

for a = 1:length(points_x)
    plotSIRDynamics(points_x(a), R0vec(z),points_y(a), 1, gamma, lambda1, delta1, lambda2, delta2, 0); % individuals adopt based on incidence
end

%%%%%%%%%%%%%%%%%%%%%%%%%%%% Figure S4 %%%%%%%%%%%%%%%%%%%%%%%%%%%%

susceptFracVec = 0:0.02:1;
R0vec = 2;
colorVec = jet(length(R0vec));
transmissionReduction = 0:0.02:1;

% SIR model parameters
gamma = 1;   % Recovery rate
lambda1 = 100; % Rate of susceptibles with lower tolerance to switch to masked class
delta1 = .1; % Rate of masked class with lower tolerance for adoption to remove intervention (mask) to return to susceptible class
lambda2 = 1; % Rate of susceptibles with higher tolerance to switch to masked class
delta2 = .1; % Rate of masked class with higher tolerance for adoption to remove intervention (mask) to return to susceptible class

recoveredFracVec = zeros(length(susceptFracVec),length(transmissionReduction),length(R0vec));
xVec = zeros(length(susceptFracVec),length(transmissionReduction),length(R0vec));
effectivenessVec = zeros(length(susceptFracVec),length(transmissionReduction),length(R0vec));

for z = 1:length(R0vec)
    for y = 1:length(transmissionReduction)
        for x = 1:length(susceptFracVec)
            recoveredFracVec(x,y,z) = plotSIRDynamics(susceptFracVec(x), R0vec(z),transmissionReduction(y), 0, gamma, lambda1, delta1, lambda2, delta2, 1); % individuals adopt based on infection level
            xVec(x,y,z) = susceptFracVec(x);
            effectivenessVec(x,y,z) = transmissionReduction(y);
        end
    end
end


figure

subplot(1,2,1)
[x_new, e_new] = meshgrid(susceptFracVec, transmissionReduction);
contourf(x_new,e_new,recoveredFracVec(:,:,z)', 'LevelStep', 0.01)
xlabel('Fraction of Population with Low Risk Tolerance', 'FontSize', 14)
ylabel('Effectiveness of Intervention (\epsilon)', 'FontSize', 14)
h = colorbar;
title(h, 'Epidemic Size', 'FontSize', 14)
title(['R_0=', num2str(R0vec(z)), ', \lambda_1=', num2str(lambda1), ', \delta_1=', num2str(delta1), ', \lambda_2=', num2str(lambda2), ', \delta_2=', num2str(delta2)]);

hold on
points_x = 0.6*ones(100,1);
points_y = rand(100,1);

scatter(points_x, points_y, 'r', 'filled')

subplot(1,2,2)

for a = 1:length(points_x)
    plotSIRDynamics(points_x(a), R0vec(z),points_y(a), 1, gamma, lambda1, delta1, lambda2, delta2, 1); % individuals adopt based on infection level
end


%%%%%%%%%%%%%%%%%%%%%%%%%%%% Figure S5 %%%%%%%%%%%%%%%%%%%%%%%%%%%%

susceptFracVec = 0:0.02:1;
R0vec = 2;
colorVec = jet(length(R0vec));
transmissionReduction = 0:0.02:1;

% SIR model parameters
gamma = 1;   % Recovery rate
lambda1 = 100; % Rate of susceptibles with lower tolerance to switch to masked class
delta1 = .1; % Rate of masked class with lower tolerance for adoption to remove intervention (mask) to return to susceptible class
lambda2 = 1; % Rate of susceptibles with higher tolerance to switch to masked class
delta2 = .1; % Rate of masked class with higher tolerance for adoption to remove intervention (mask) to return to susceptible class

recoveredFracVec = zeros(length(susceptFracVec),length(transmissionReduction),length(R0vec));
xVec = zeros(length(susceptFracVec),length(transmissionReduction),length(R0vec));
effectivenessVec = zeros(length(susceptFracVec),length(transmissionReduction),length(R0vec));

for z = 1:length(R0vec)
    for y = 1:length(transmissionReduction)
        for x = 1:length(susceptFracVec)
            recoveredFracVec(x,y,z) = plotSIRDynamics(susceptFracVec(x), R0vec(z),transmissionReduction(y), 0, gamma, lambda1, delta1, lambda2, delta2, 1); % individuals adopt based on infection level
            xVec(x,y,z) = susceptFracVec(x);
            effectivenessVec(x,y,z) = transmissionReduction(y);
        end
    end
end


figure

subplot(1,2,1)
[x_new, e_new] = meshgrid(susceptFracVec, transmissionReduction);
contourf(x_new,e_new,recoveredFracVec(:,:,z)', 'LevelStep', 0.01)
xlabel('Fraction of Population with Low Risk Tolerance', 'FontSize', 14)
ylabel('Effectiveness of Intervention (\epsilon)', 'FontSize', 14)
h = colorbar;
title(h, 'Epidemic Size', 'FontSize', 14)
title(['R_0=', num2str(R0vec(z)), ', \lambda_1=', num2str(lambda1), ', \delta_1=', num2str(delta1), ', \lambda_2=', num2str(lambda2), ', \delta_2=', num2str(delta2)]);

hold on
points_x = rand(100,1);
points_y = 0.3*ones(100,1);

scatter(points_x, points_y, 'r', 'filled')

subplot(1,2,2)

for a = 1:length(points_x)
    plotSIRDynamics(points_x(a), R0vec(z),points_y(a), 1, gamma, lambda1, delta1, lambda2, delta2, 1); % individuals adopt based on infection level
end



%%%%%%%%%%%%%%%%%%%%%%%%%%%% Figure S6 %%%%%%%%%%%%%%%%%%%%%%%%%%%%
% Repeat this script with corresponding parameters to generate each panel

figure
susceptFracVec = 0.2;
R0vec = 2; 
colorVec = jet(length(R0vec));
transmissionReduction = .7;

% SIR model parameters
cVec = [0.01:0.05:1];
lambdaAvg = [0.5, 1:5:100];
gamma = 1;   % Recovery rate
delta1 = .5; % Rate of masked class with lower tolerance for adoption to remove intervention (mask) to return to susceptible class
delta2 = .5; % Rate of masked class with higher tolerance for adoption to remove intervention (mask) to return to susceptible class

recoveredFracVec = zeros(length(cVec),length(lambdaAvg));

for x = 1:length(cVec)
    for y = 1:length(lambdaAvg)
        recoveredFracVec(x,y) = plotSIRDynamics(susceptFracVec, R0vec, transmissionReduction, 0, gamma, lambdaAvg(y)*(1-susceptFracVec*cVec(x))/(1-susceptFracVec), delta1, cVec(x)*lambdaAvg(y), delta2, 1); %Arithmetic parameterization
    end
end

[c_new, l_new] = meshgrid(cVec, lambdaAvg);
contourf(c_new,l_new,recoveredFracVec', 'LevelStep', 0.01)
xlabel('\lambda Homogeneity Index', 'FontSize', 14)
ylabel('\lambda_{Arithmetic Average}', 'FontSize', 14)
h = colorbar;
title(h, 'Epidemic Size', 'FontSize', 14)
title(['R_0=', num2str(R0vec), ', x_1=', num2str(susceptFracVec), ', \epsilon=', num2str(transmissionReduction), ', \delta_1=', num2str(delta1),', \delta_2=', num2str(delta2)]);

\end{lstlisting}

\end{document}